\documentclass[11pt,preprint]{aastex}
\usepackage{natbib}
\usepackage{graphicx}
\usepackage{longtable}
\usepackage{graphics}

\shorttitle{Variability in Cyg OB2}
\shortauthors{Henderson et al.}

\begin{document}

\title{An \textit{R}- and \textit{I}-Band Photometric Variability Survey of the Cygnus OB2 Association\altaffilmark{1}}
\author{
{C.~B. Henderson}\altaffilmark{2},
{K.~Z. Stanek}\altaffilmark{2},
{O. Pejcha}\altaffilmark{2}, and
{J.~L. Prieto}\altaffilmark{3}
}
\altaffiltext{1}{Based on observations obtained using the 1.3m McGraw-Hill telescope at the MDM Observatory}
\altaffiltext{2}{Department of Astronomy, The Ohio State University, 140 W. 18th Ave., Columbus, OH 43210, USA}
\altaffiltext{3}{Hubble and Carnegie-Princeton Fellow at Carnegie Observatories, 813 Santa Barbara St., Pasadena, CA 91101, USA}
\email{henderson@astronomy.ohio-state.edu}

% -----------------------------------------------------------------
\begin{abstract}

	We present a catalog of photometrically variable stars discovered within two $21.3\arcmin \times 21.3\arcmin$ fields centered on the Cygnus OB2 association.  There have hitherto been no deep optical variability studies of Cyg OB2 despite it being replete with early-type massive stars, perhaps due to the high and variable extinction (up to $A_{V}$~$\approx$~20) that permeates much of the region.  Here we provide results of the first variability study with this combination of spatial coverage ($\sim$~0.5 deg) and photometric depth ($R$~$\sim$~21 mag).  We find 121 stars to be variable in both \textit{R}- and \textit{I}-band, 116 of them newly discovered.  Of the 121 variables, we identify 27 eclipsing binaries (EBs) and eclipsing binary candidates, 20 potential Herbig Ae/Be stars, and 52 pulsating variables.  Confirming both the status and the cluster membership of the Herbig Ae/Be stars would address the uncertainty regarding the age and star formation history of Cyg OB2.  We match our catalog to known variables and binaries in the region, 2MASS near-IR (NIR) data, and \textit{Chandra} X-ray observations to find counterparts to new variables in other wavelengths.

\end{abstract}
% -----------------------------------------------------------------

\keywords{Cyg OB2 --- binaries: eclipsing --- stars: massive --- stars: variables: general}

%%%%%%%%%%%%%%%%%%%%%%%%%%%%%%%%%%%%%%%%%%%%%%%%%%%%%%%%%%%%%%%%%%%%%%%%%%%%%%%
\section{Introduction}
%%%%%%%%%%%%%%%%%%%%%%%%%%%%%%%%%%%%%%%%%%%%%%%%%%%%%%%%%%%%%%%%%%%%%%%%%%%%%%%

	Procuring accurate fundamental measurements of the masses and radii of stars remains one of the most outstanding questions in astronomy.  This is crucial in the high-mass regime, which is essential for determining the slope of the initial mass function (IMF) for high-mass stars.  The sensitivity of stellar models to the initial mass cannot be overstated.  This, then, necessitates fundamental determinations of stellar masses with errors $\le$1--3$\%$ \citep{torres10} in order to reject models with incorrect underlying physics.  As emphasized in the review by \citet{torres10}, there are only 95 detached binary systems of any mass in which the mass and radius of both components are known to $\pm3\%$ or better.  All but one are eclipsing binaries, with the sole exception being the Sun's nearest neighbor, $\alpha$ Centauri.  Of the 95 total systems there are only eight with $M>15~M_{\odot}$, three with $M>20~M_{\odot}$, and none with $M>30~M_{\odot}$.  Even when applying a less stringent uncertainty cutoff, only 14 stars with $M>30~M_{\odot}$ have both their mass and radius known to $\pm10\%$ or better \citep{bonanos09}.  This paucity of accurately determined stellar masses and radii at the high-mass end requires an expedient redressing.  Fundamental measurements of properties for a statistical sample of massive stars will allow us to constrain theoretical models, anchor the IMF at the high-mass end, and better comprehend stellar structure and evolution.

	The Cygnus OB2 association presents the ideal laboratory within which to learn about the fundamental properties of massive stars.  It is the nearest massive star-forming region, making it a prime target observationally, even in spite of the extinction in the larger Cygnus region, which can range up to $A_{V}\approx20$.  Furthermore, it is known to be replete with an estimated 2,600~$\pm$~400 OB stars, including 120~$\pm$~20 O-type stars, and have a total inferred mass of $\sim(4-10)\times10^{4}~M_{\odot}$ \citep{knodlseder00}.  The region has been the focus of several studies at a variety of wavelengths, dating back to \citet{munch53}, who initially  noticed a clustering of 11 blue giants near $\gamma$ Cygni.  \citet{johnson54} were the first to obtain accurate spectral types, finding 10 to be O-type and one to be an early B-type star.  The subsequent discovery of the heavily reddened and extremely luminous star Cyg OB $\#$12 \citep{morgan54a} was followed by seven \citep{morgan54b} and then 12 additional early-type stars \citep{schulte56a}.  The first comprehensive study of the morphology of the association estimated more than 3,000 total members, of which at least 300 are early-type (OB) stars, and a total mass of $\sim(0.6-2.7)\times10^{4}~M_{\odot}$ \citep{reddish66}.

	The first synoptic optical study of VI Cygnus produced \textit{UBV} photometry of 799 stars in the region and obtained spectral types for 76 stars, most of which are the brightest and bluest members of the association \citep{massey91}.  They found a distance of 1.7 kpc and the slope of the IMF to be $\Gamma=-1.0\pm0.1$.  This provided the first photometric catalog of Cyg OB2 and spectroscopically identified the largest number of OB association members.  \citet{pigulski98} performed the first variability survey of the region, obtaining \textit{I}-band photometry of the central 6 $\times$ 12 arcmin$^{2}$.  They found 29 variables, five of which are O-type, and at least five eclipsing binaries.  \citet{knodlseder00} studied the region using 2MASS data to characterize the stellar population.  They aver an IMF slope of $\Gamma=-1.6\pm0.1$ and assert that the large total mass, high central density, and overall compactness of the region require that it be classified as a young globular cluster, a claim originally made by \citet{reddish66}.

	Precise classification notwithstanding, early-type members continued to be discovered spectroscopically in the region.  \citet{comeron02} used 2MASS photometry to identify hundreds of potential early-type stars and spectroscopically found 77 of them to be early-type members.  Of those, 31 had prior classifications, 24 of which are O-type and the remaining seven B-type, causing them to extrapolate that a total of 60 of their sample of 77 are O-type stars and that Cyg OB2 contains between 90--100 total O-type stars.  \citet{hanson03} took spectra of 14 of the remaining 46 and identified all 14 to be early-type, perhaps indicating that the remaining 31 (one was discovered to have been previously studied) are also early-type.  \citet{hanson03} then extrapolated the locally derived cluster luminosity function to argue that there could be tens or even one hundred similar massive open clusters within the Milky Way.  Recently, \citet{kiminki07} built off the canonical optical study of \citet{massey91}, spectroscopically following up 183 of their photometrically observed Cyg OB2 members.  Seventy-three possessed extant spectral classifications as OB stars and 73 of the remaining 110 were also found to be OB stars.  Among these they flagged 36 probable and nine possible single-lined spectroscopic binaries (SB1s) as well as three new and eight candidate double-lined spectroscopic binaries (SB2s).

	Several additional binaries have been discovered in the Cyg OB2 region in the past few years.  \citet{kiminki08} presented orbital solutions for two SB1s and one SB2 that they had identified as SB-candidates in their earlier work.  Cygnus OB2 $\#9$, a non-thermal radio-emitter, was confirmed as a binary \citep{naze08}, supporting the theoretical wind-wind collision scenario.  Two EBs in the Cyg OB2 region were found via data in the NSVS survey \citep{otero08a, otero08b, stroud10}.  \citet{kiminki09} determined the orbital parameters of four additional OB SBs.  This brings the total number of known OB binary systems with full photometric and/or spectroscopic solutions within Cyg OB2 to 18.

	Furthermore, the uncertainty regarding the age and star-formation history of Cyg OB2 makes this cluster all the more intriguing.  The abundance of heavily-reddened early-type stars found by \citet{massey91} initially suggested a single epoch of star formation and a corresponding age of 1--3 Myr.  Several of the early-type candidates identified by \citet{comeron02} were confirmed as OB stars by \citet{hanson03}, but their dubious status as cluster members casts doubt on the 1--1.5 Myr age that their membership would necessitate.  \citet{hanson03} goes on to question the membership status of the remaining stars in the \citet{comeron02} sample, and the presence of evolved O-stars such as the supergiant Cyg OB2 $\#$9 requires an age of at least 3 Myr.  More recently, IPHAS observations by \citet{drew08} revealed a clustering of $\sim$200 A-type dwarfs primarily to the south of Cyg OB2 and within the 5--7 Myr age range.  \citet{comeron08} used the same method of \citet{comeron02} to investigate an annular region between 1 and 2 degrees whose center lies slightly to the north of Cyg OB2.  They uncovered a diffuse spread of evolved early-type stars wtih ages up to $\sim$10 Myr that they identify as likely being field population stars.  Finally, \citet{wright10} used isochrone fits to NIR color-magnitude diagrams (CMDs) to derive ages of 3.5 and 5.25 Myr for their two fields.  Given the large spread in ages, it is unclear whether the central core of Cyg OB2 is merely the latest in a series of epochs of star formations, if it is the product of a parent cluster, or if the populations of older and more evolved stars are simply not cluster members.

	The Cygnus OB2 association is also interesting as a source of high-energy ($\geq$MeV) emission.  Due to limitations of angular resolution, most observations thus far cannot clearly separate Cyg OB2 from nearby parts of the Cygnus region more generally.  This region is bright in gamma rays from both point sources and diffuse emission, ranging from the MeV to the GeV to the TeV \citep{hartman99, prodanovic07, abdo07, albert08, romero10}.  Neutrinos and gamma rays from this region will be important for testing models of cosmic ray production \citep{anchordoqui07, beacom07, halzen07, anchordoqui09}.  With present and future instruments, it should be possible to better resolve the young Cyg OB2, which would allow the isolation of the high-energy emission from pre-explosion massive stars as opposed to post-explosion remnants.

	This paper describes the findings of the first deep optical two-band (\textit{R}$_{C}$, \textit{I}$_{C}$) photometric variability survey of the Cygnus OB2 region.  We find 121 variables and provide variable classifications for them.  In $\S2$ we give a full description of the photometric observations.  The details of the photometric calibration and light curve production are in $\S3$.  We present the results of the variability search in $\S4$ and discuss the different variability classes separately.  We match our variability catalog to the 2MASS Point Source Catalog in order to find NIR counterparts and then to a recent X-ray survey of Cyg OB2 \citep{wright09}, both of which are discussed in $\S5$.  Finally, $\S6$ summarizes the results of the survey and future directions of the project.

%%%%%%%%%%%%%%%%%%%%%%%%%%%%%%%%%%%%%%%%%%%%%%%%%%%%%%%%%%%%%%%%%%%%%%%%%%%%%%%
\section{Data}
\label{sec:data}
%%%%%%%%%%%%%%%%%%%%%%%%%%%%%%%%%%%%%%%%%%%%%%%%%%%%%%%%%%%%%%%%%%%%%%%%%%%%%%%

% -------------------------
\subsection{Observations}
\label{sec:obs}
% -------------------------

\begin{figure}
\plotone{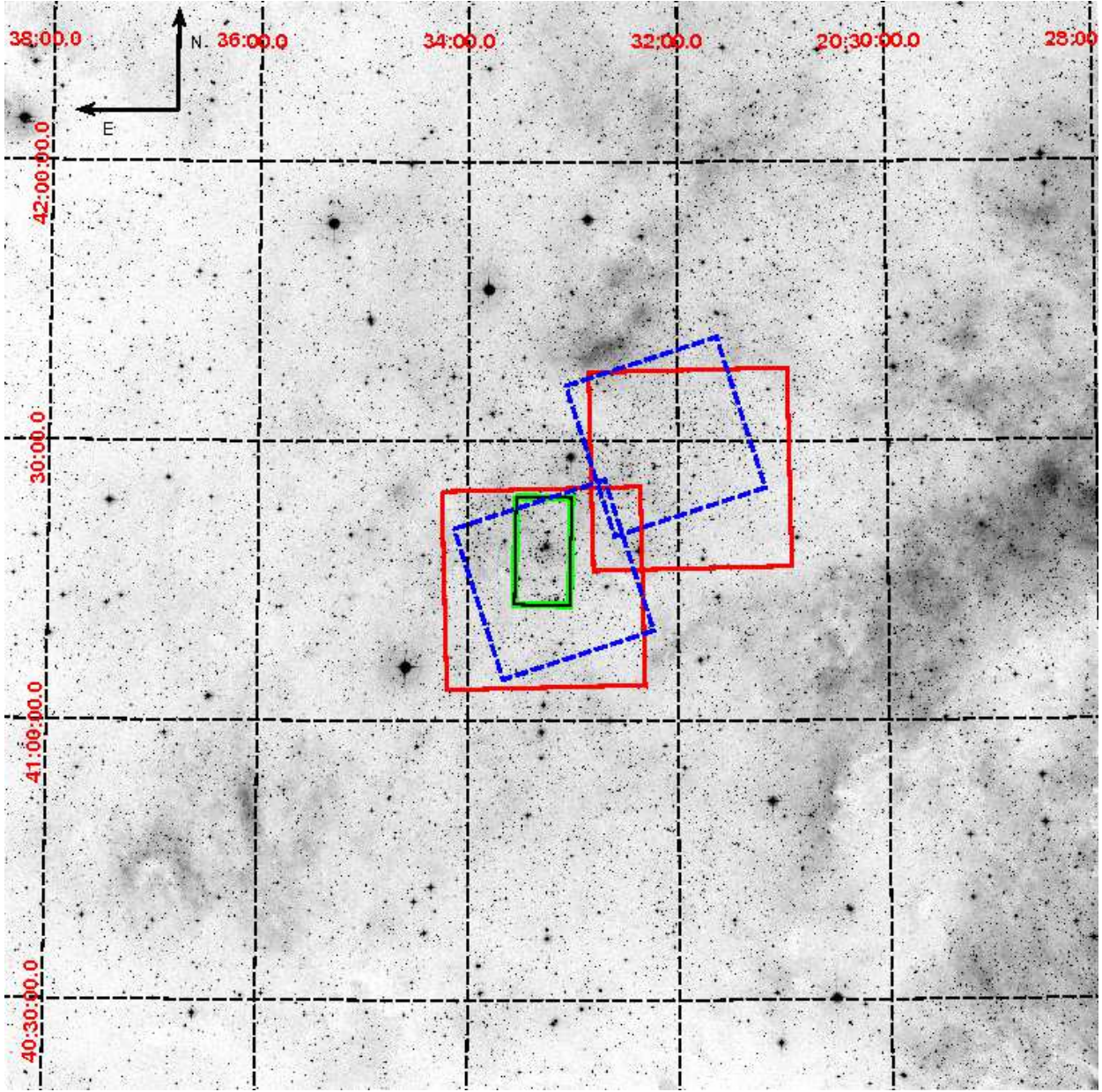}
\caption{A red filter DSS image of the Cygnus OB2 region.  The image size is $2\times2$ deg$^{2}$ and centered on (R.A., Dec)~=~(20:33:12, +41:19:00) with north up and east to the left.  The area covered by this study is shown in red while the variability study of \citet{pigulski98} is shown in green and the X-ray study of \citet{wright09} is shown in dashed blue.
}
\label{fig:fichart}
\end{figure}

	The data were obtained on the MDM 1.3m telescope using the 4K imager, which has a field of view (FoV) of $21.3\arcmin \times 21.3\arcmin$ and a pixel scale of $0.315\arcsec$ per pixel.  The observations consist of 60s exposures taken through the Cousins \textit{R}-filter and 10s exposures taken through the Cousins \textit{I}-filter.  The data were collected over 19 nights during the fall of 2007, spanning September 12 - October 24, and 18 nights during the summer of 2009, spanning May 30 - June 25.  Two pointings were used to provide greater spatial coverage of the Cygnus OB2 region.  The first field was centered at ($\alpha = 20^\mathrm{h}33^\mathrm{m}17^\mathrm{s}, \hspace{0.5pc} \delta = +41^\mathrm{d}14^\mathrm{m}15^\mathrm{s}$) and the second field to the north at ($\alpha = 20^\mathrm{h}31^\mathrm{m}53^\mathrm{s}, \hspace{0.5pc} \delta = +41^\mathrm{d}27^\mathrm{m}00^\mathrm{s}$), both chosen in order to cover the central core of the cluster and coincide with the study of \citet{massey91}.  Table~\ref{obs} gives an overview of the observations.  Figure~\ref{fig:fichart} shows a 2 $\times$ 2 deg$^{2}$ image of the Cygnus OB2 region with overlays of this study, the previous variability study of \citet{pigulski98}, and the X-ray survey of \citet{wright09}.

% -----------------------------
\subsection{Basic Data Reduction}
\label{sec:reduc}
% -----------------------------

	We apply basic CCD reductions using a custom software pipeline\footnote[1]{http://www.astronomy.ohio-state.edu/$\sim$jdeast/4k/proc4k.pro} developed by Jason Eastman.  The routine performs overscan subtraction, trims the images, and uses dome flats to flat field the data.  The 4K chip suffers from crosstalk contamination, a subtle but insidious effect, and so the procedure was modified to properly correct for this.  Crosstalk results from the chip being divided into four quadrants of equal pixel size that are read out by four different amplifiers simultaneously.  During the readout phase, signal from bright stars being read out on a given quadrant, referred to as the source quadrant, will reverberate through the electronics and leave residual signal on the other three quadrants, called victim quadrants.  The source signal is reflected across the three axes of symmetry of the chip, making the contamination predictable and predominantly tractable.

	Before the residual signal can be removed, a set of crosstalk coefficients, which specify the fraction of the source signal that is reflected onto each victim quadrant, must be calculated.  The coefficients are small enough ($\sim0.04\%-0.1\%$) that the problem can be sufficiently dealt with via a simple solution.  Using a set of a few hundred images of our data from both the 2007 and 2009 seasons, we find that the coefficients are not only constant in the count regime 45000 $\leq$ counts $<$ 65535, but also stable across the two-year baseline.  Information cannot be backed out of reflected signals due to saturated (65535 counts on the 4K imager) sources, so pixels contaminated by saturated stars are set to zero, removed from the light curves, and otherwise ignored.

%%%%%%%%%%%%%%%%%%%%%%%%%%%%%%%%%%%%%%%%%%%%%%%%%%%%%%%%%%%%%%%%%%%%%%%%%%%%%%%
\section{Photometric Calibration and Light Curve Production}
\label{sec:lightcurves}
%%%%%%%%%%%%%%%%%%%%%%%%%%%%%%%%%%%%%%%%%%%%%%%%%%%%%%%%%%%%%%%%%%%%%%%%%%%%%%%

	We use astrometry.net \citep{lang10} to do astrometry and we extract photometry using the ISIS package \citep{alard98, alard00} in conjunction with the DAOPHOT/ALLSTAR package \citep{stetson87, stetson92}, the procedural details of which are discussed in \citet{mochejska03} and \citet{hartman04}.  We did not observe standard stars, so this requires that we match to other calibrated catalogs and apply a computed magnitude offset.  For the \textit{R}-band data we match our full catalog to an overlapping region of stars observed by the IPHAS survey \citep{drew05}.  We uniquely match six stars from their catalog to ours using a matching radius of 1$\arcsec$.  We then apply the transformation equation given by Lupton et al. (2005)\footnote[2]{http://www.sdss.org/dr6/algorithms/sdssUBVRITransform.html},
	\begin{equation}                                                                                                                                   R = r - 0.2936\cdot(r - i) - 0.1439,                                                                                                   \end{equation}                                                                                                                                    to transform the IPHAS SDSS magnitudes into Cousins \textit{R}-band magnitudes.  From this we derive the magnitude offset and calibrate our \textit{R}-band data.  To calibrate our \textit{I}-band data with the catalog of \citet{pigulski98} we use a more stringent star selection, as there is more spatial overlap between the two data sets.  For a star to be used for the calibration, it has to satisfy the following criteria: 1) have a unique match within 1$\arcsec$, 2) have no additional matches within 5$\arcsec$, 3) not be classified by them as a variable, and 4) not be saturated or blended in our own database.  This ultimately leaves 133 stars that we then use to calibrate our \textit{I}-band data.  As each of our fields has a different reference frame, there is a slight magnitude offset between fields that can be calculated using stars in the overlap region.

\begin{figure}
\plotone{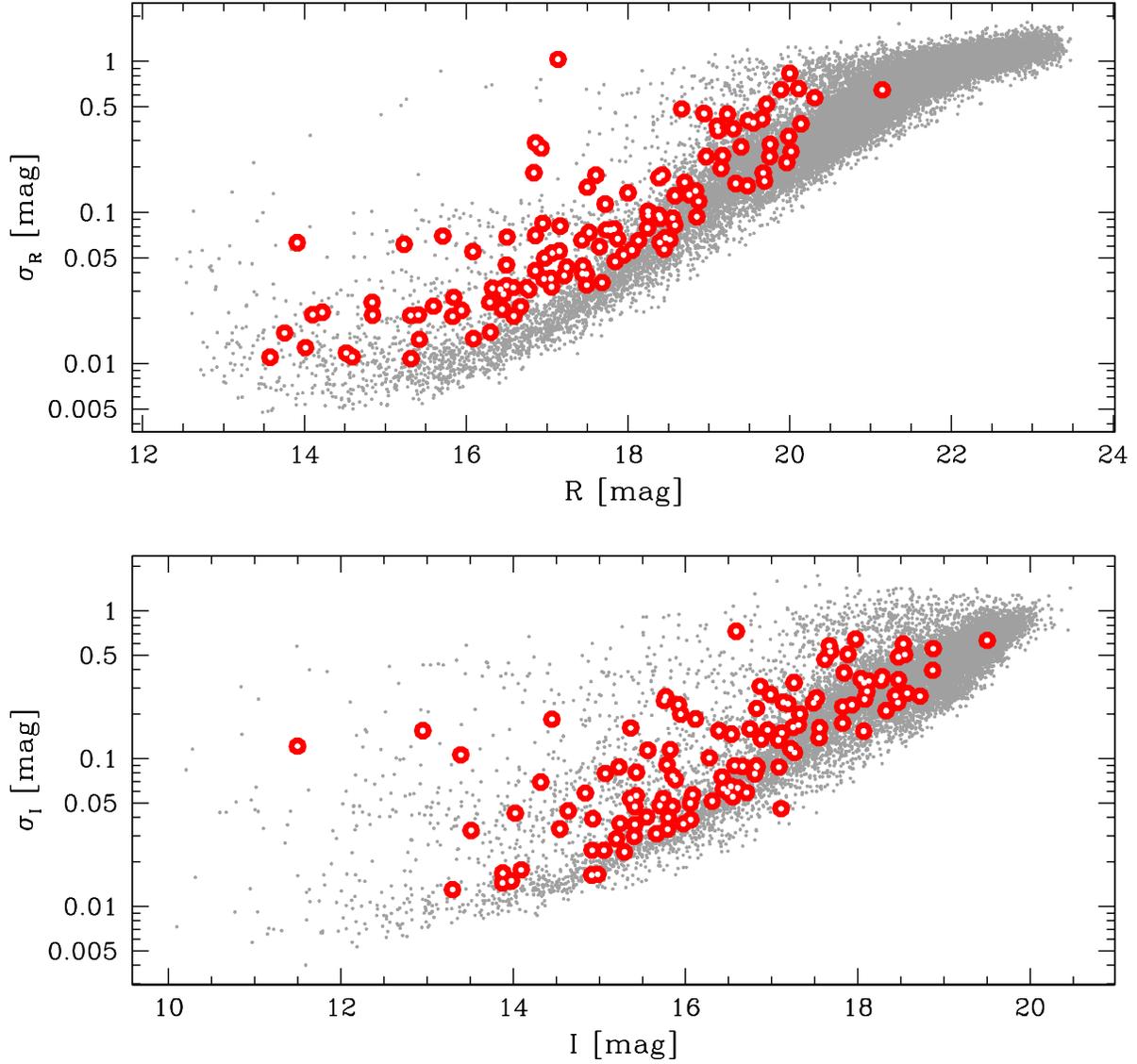}
\caption{The rms as a function of magnitude for the 23,561 $R$-band light curves (top panel) and 12,611 $I$-band light curves (bottom panel).  The 121 stars that display both $R$- and $I$-band photometric variability are shown in red.}
\label{fig:sigma}
\end{figure}

	We perform a variability search using the VARTOOLS light curve analysis program \citep{hartman08}.  We employ both Analysis of Variation (AoV) \citep{schwarzenberg-czerny89, devor05} and Lomb-Scargle (LS) \citep{lomb76, scargle82, press89, press92} period search algorithms.  We then inspect light curves with high significance by eye to find periodic variables and use the Stetson variability index \citep{welch93} to select irregular variable stars.  Figure~\ref{fig:sigma} shows the light curve rms as a function of reference magnitude for our 23,561 $R$-band light curves and 12,611 $I$-band light curves.  We find variables as deep as $R\sim21$ and $I\sim19.5$ and go fainter in $R$ due to the longer exposure time.  Many stars with high rms are not listed as variable because we establish the criterion that stars must exhibit simultaneous two-band variability, which eliminates false positives and mitigates any potential residual effects of crosstalk contamination from saturated sources.  As described in \citet{stetson96}, stars displaying variability in only one band will have a variability index near zero.  Only stars that exhibited clear two-band variability are selected, 121 in total.

%%%%%%%%%%%%%%%%%%%%%%%%%%%%%%%%%%%%%%%%%%%%%%%%%%%%%%%%%%%%%%%%%%%%%%%%%%%%%%%
\section{Variable Stars}
\label{sec:varstars}
%%%%%%%%%%%%%%%%%%%%%%%%%%%%%%%%%%%%%%%%%%%%%%%%%%%%%%%%%%%%%%%%%%%%%%%%%%%%%%%

	Prior to discussing the parameters of the variables, the various naming conventions must be explained.  \citet{munch53} numbered the first 11 stars discovered in Cyg OB2, a system that \citet{morgan54a} expanded for star number 12, \citet{morgan54b} for stars 13--19, and \citet{schulte56a, schulte56b} for stars 20--85.  We follow the convention of the literature, which is to precede these stars with \textquoteleft Schulte\textquoteright.  All variable stars in the GCVS are identified as such.  The primary numbering system is that of \citet{massey91}, for which the designation MT91 precedes star ID numbers, from 1--799.  The variability study of \citet{pigulski98} extends this system to 975, and is referred to as MTE (MT91 Extended).  The designations A36 and A45 (GSC 03161--00815, 2MASS J20294666+4105083, respectively) arise from the convention introduced by \citet{comeron02} and followed by \citet{kiminki09}.  Stars within our own variability catalog are given a VarID, and numbers increase from brightest to faintest mean $R$-band magnitude.

	\citet{pigulski98} performed the first variability survey of the region, detecting 455 stars in \textit{I}-band and finding 29 of them to be variable, including five EBs and five O-type stars.  Our study goes much deeper, covers significantly more area, and furthermore provides a two-band photometric catalog.  We identify 121 stars that exhibit variability in both \textit{R}- and \textit{I}-band.  Of these, 10 are EBs, nine of them newly discovered, 17 are EB-candidates, 20 are potential Herbig Ae/Be stars, and 52 are pulsating stars.  We first discuss the results of matching our catalog to the literature.  Then we discuss each variability class individually: first the EBs and EB-candidates, then the pulsating variables, and finally the potential Herbig Ae/Be stars.

% -----------------------------
\subsection{Literature Comparison}
\label{sec:lit}
% -----------------------------

	We find 121 stars to be variable in both \textit{R}- and \textit{I}-band within the Cygnus OB2 association.  After doing a literature search, we match our catalog to all previously known variable stars in Cyg OB2, listed in Table~\ref{photvar}, using a matching radius of 1$\arcsec$.  Of the 30 known variable stars in Cyg OB2, we recover variability in at least one band for nine of them.  Seven stars are without matches while the remaining 14 are either saturated, photometrically constant, or outside our FoV.  We discuss those with matches here.

	V2186 Cyg is an EB of the Algol type with a period of 2.9788 days \citep{pigulski98}.  Though it is saturated in our \textit{R}-band data, we match to it in \textit{I}-band and successfully recover the period.  MT91 460 is an irregular variable with an \textit{I}-band amplitude of 0.28 mag \citep{pigulski98}.  We do not match to it in our \textit{R}-band data, but find it to exhibit a 0.7 mag drop in \textit{I}-band, with the overall shape of the light curve resembling that of a long period variable.  Schulte 57, VarID 020 in our catalog, was previously classified as a variable of unknown type with a 0.728 day period.  We recover the period in both bands and classify it as a pulsating star.
	
	The remainder of the stars discussed here are part of the MTE system, introduced by \citet{pigulski98} to extend the catalog from 799 to 975 stars.  They report MTE 831 to be irregular with an \textit{I}-band amplitude of 0.22 mag.  It matches to our VarID 050, with a 10 day period and a 0.3 mag amplitude in both \textit{R}- and \textit{I}-band.  MTE 849 (VarID 021) is an EB of the Algol type.  We report a period of 2.89 days, the first known for this system.  MTE 876 (VarID 036), previously a variable of unknown type, appears to have a characteristically pulsating light curve, with a period of 1.53 days and a 0.1 mag amplitude.  We agree with their classification of MTE 900 (VarID 038), which varies erratically with an amplitude of 1.0 mag on long timescales, perhaps indicating that it is an irregular or long-period variable.  MTE 916 (VarID 011) similarly appears to be a long-period variable, though with a smaller 0.3 mag amplitude.  \citet{pigulski98} classify MTE 973 as a potential EB, and we find it to be constant in \textit{R}-band but to have a 0.2 amplitude and a 0.88 day period in \textit{I}-band, though with no obvious eclipsing signature.

% -----------------------------
\subsection{Binary Systems}
\label{sec:bins}
% -----------------------------

\begin{figure}
\plotone{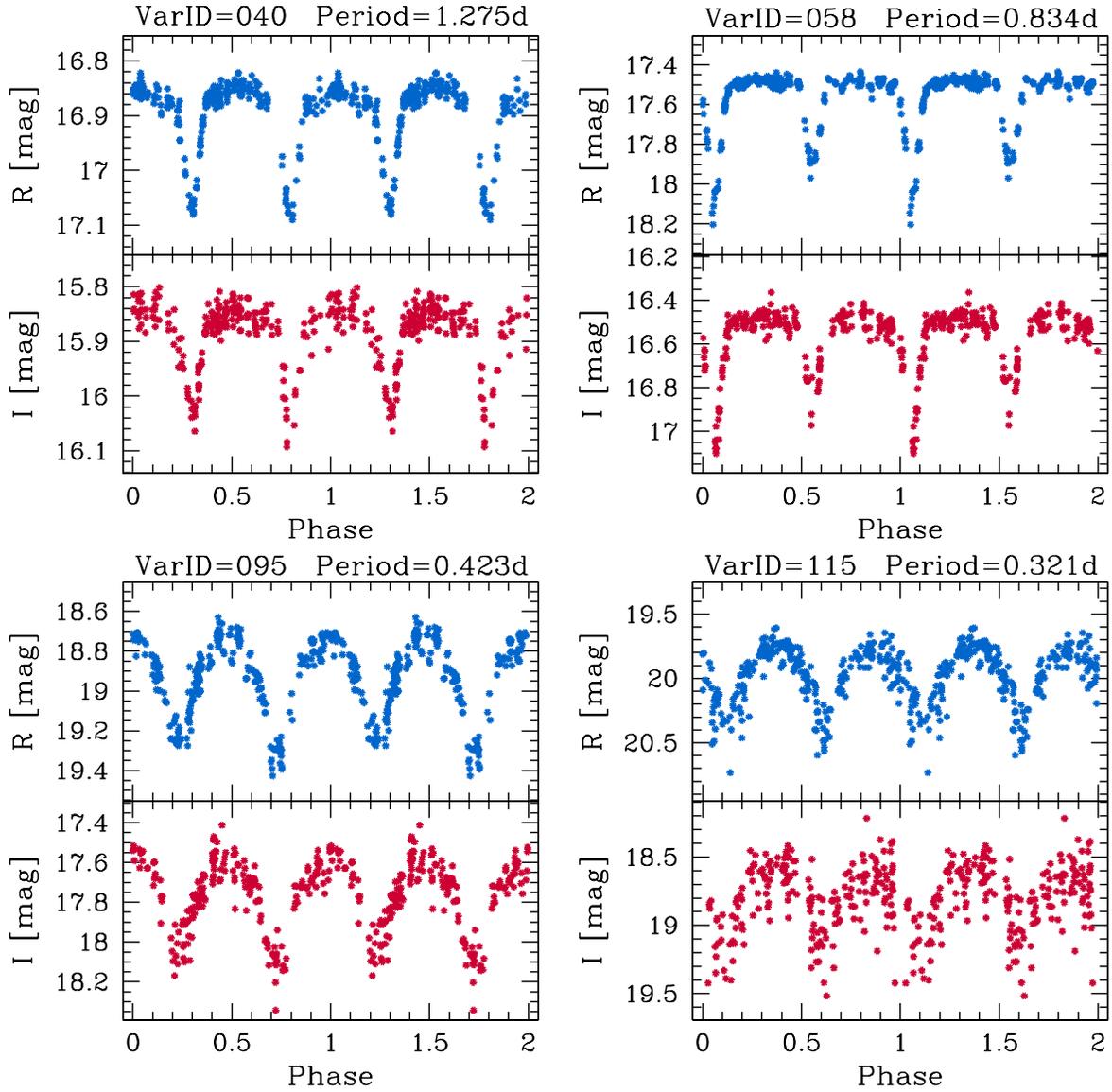}
\caption{Two-band light curves of four EBs spanning a representative magnitude range.}
\label{fig:lc.ebs}
\end{figure}

\begin{figure}
\plotone{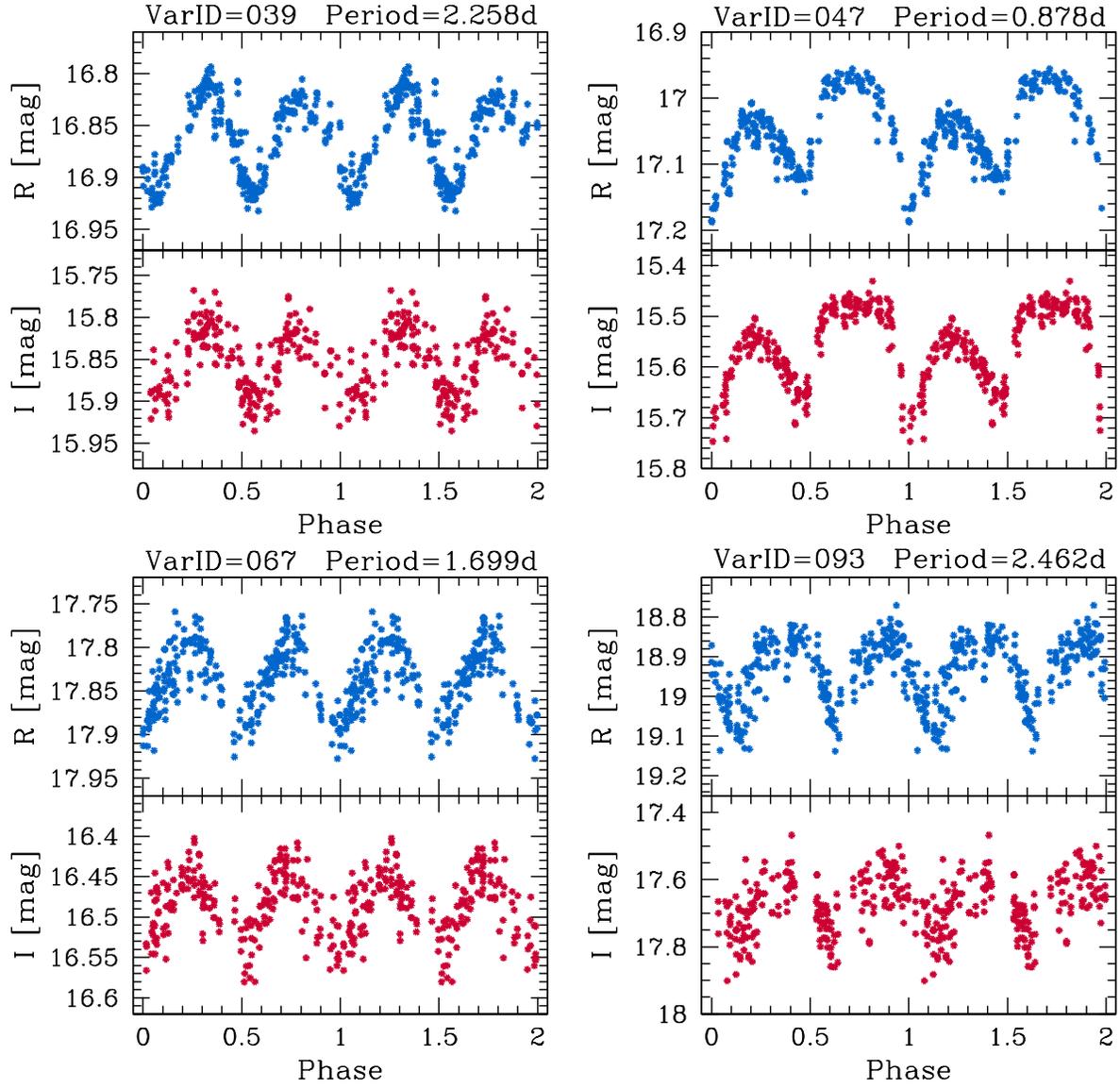}
\caption{Two-band light curves of four EB-candidates.}
\label{fig:lc.ebcands}
\end{figure}

	In addition to the eight previously known EBs within Cyg OB2, there are 13 OB SBs in Cyg OB2, four of which are single-lined and 11 of which are double-lined.  It should be noted that a few of these spectroscopic systems are also definite or possible eclipsing systems.  Those that are confirmed EBs, Schulte 27 and A36, are treated here as photometric variables.  Those that are unconfirmed, MT372 and Schulte 3, are included here only as SBs.  Of the 121 stars we identify as photometrically variable, 10 are EBs and 17 are EB-candidates.  Figure~\ref{fig:lc.ebs} shows sample light curves for four of the EBs and figure~\ref{fig:lc.ebcands} shows four of the EB-candidates.

	To determine if any of our variable stars are SBs, we match to all confirmed and potential SBs found in the literature, most of which are from the catalog of \citet{kiminki07}.  They present spectroscopic data for 146 OB stars within Cyg OB2, all initially discovered by \citet{massey91}, and find 73 new OB stars.  Restricting themselves to the 120 stars with the best data, they identify 36 probable and nine possible SB1s and three new and eight possible SB2s.  Table~\ref{specbin} lists confirmed and potential SBs within Cyg OB2.  Given that the catalog of \citet{kiminki07} reaches a limiting magnitude of $V\approx15$ and our upper limit is $R\approx13.5$, the vast majority of the stars are saturated in our catalog, and thus we find no matches.

% -----------------------------
\subsection{Pulsating Variables}
\label{sec:pulse}
% -----------------------------

\begin{figure}
\plotone{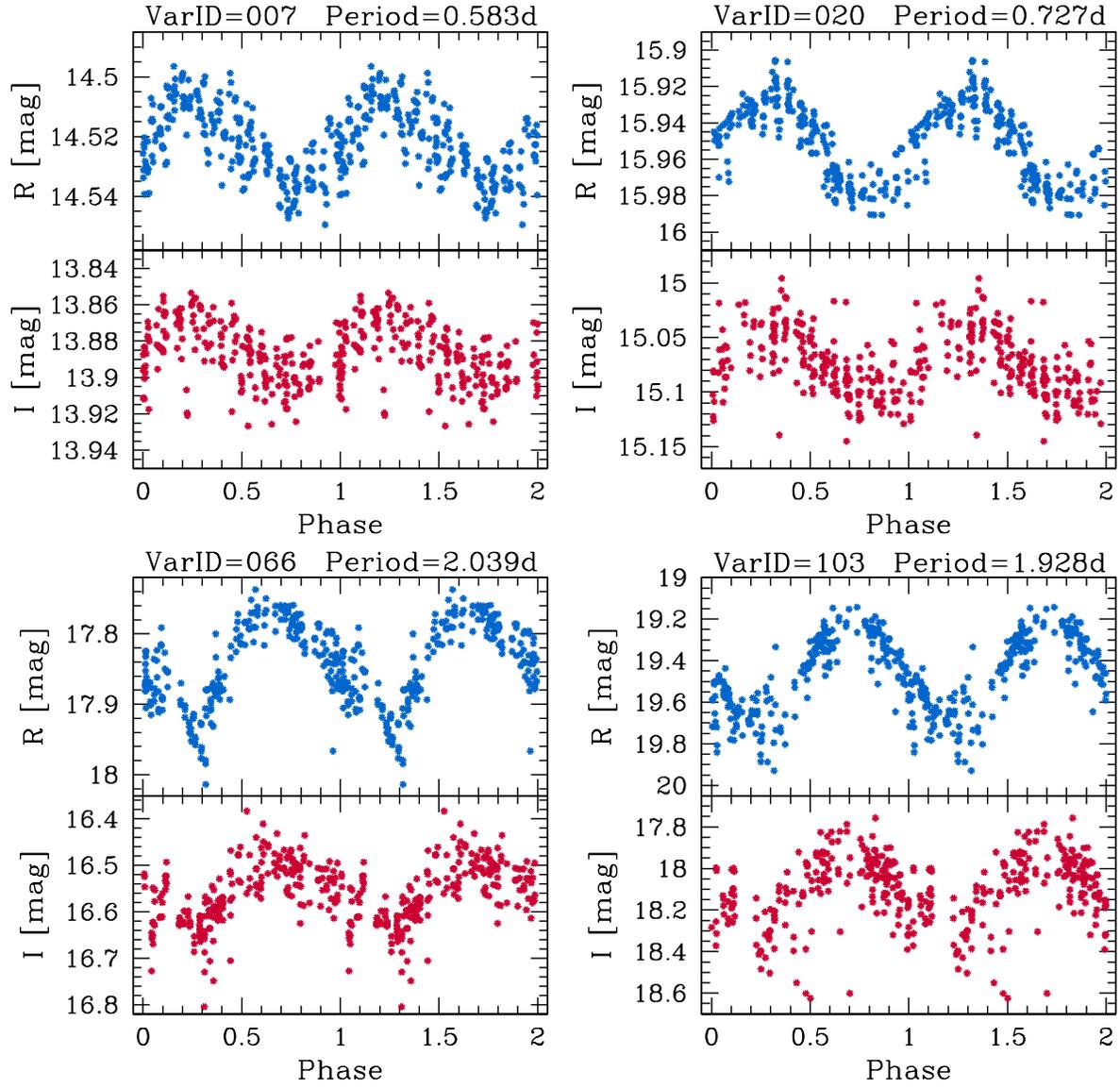}
\caption{Two-band light curves of four pulsating variables.}
\label{fig:lc.pulse}
\end{figure}

\begin{figure}
\plotone{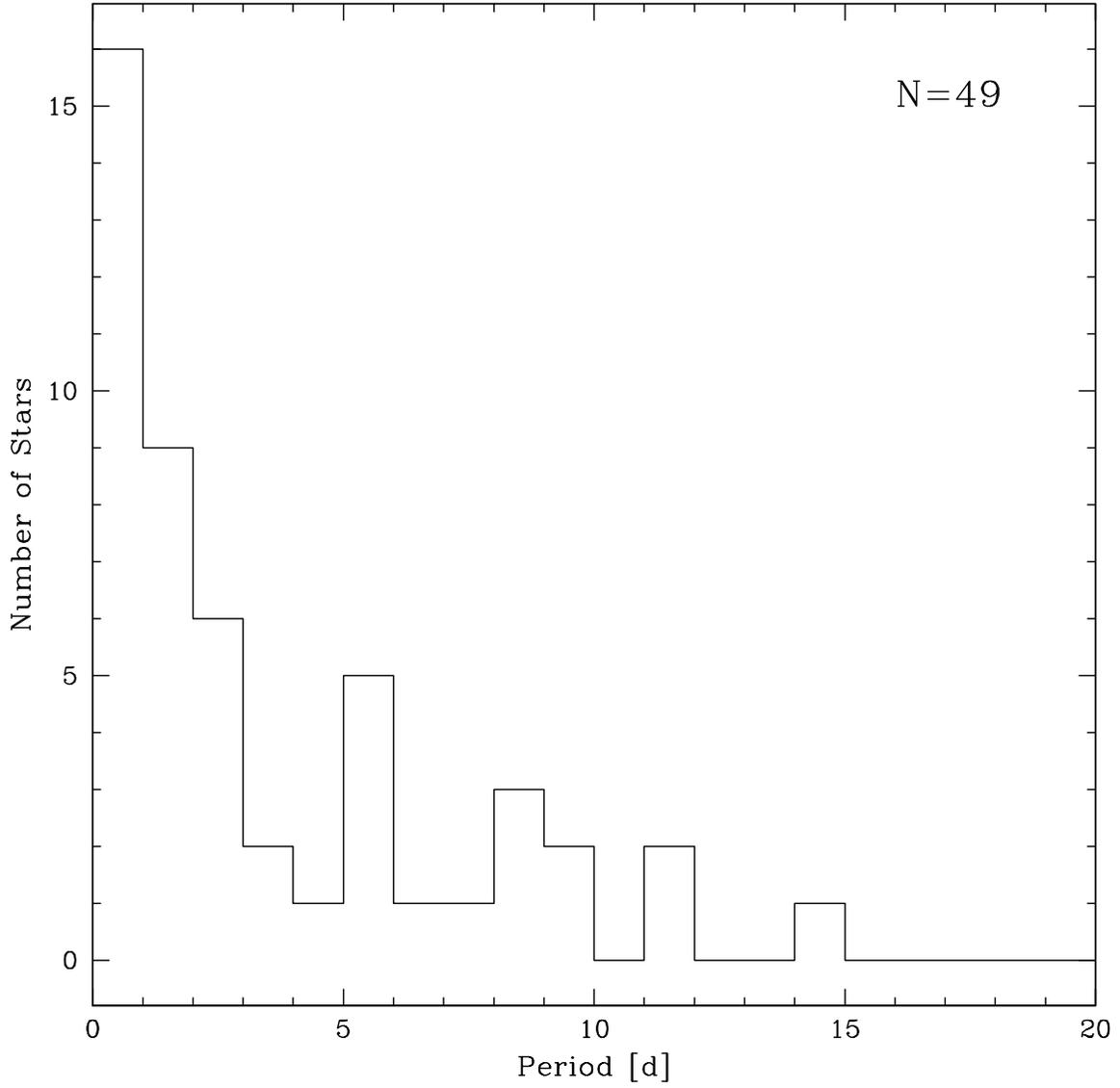}
\caption{A histogram of the period distribution of the 49 pulsating variables with periods $\leq$20 days.  Not shown are three remaining pulsating variables that have periods of $\approx$22, 27, and 59 days.}
\label{fig:perhist}
\end{figure}

	We find 52 of our stars to be pulsating variables, with $R$-band amplitudes ranging from 0.1 to 0.6 magnitudes.  Figure~\ref{fig:lc.pulse} shows sample light curves of four of the pulsating variables.  The periods range from $\sim$0.05 to $\sim$60 days.  A histogram of the period distribution from 0 to 20 days is shown in figure~\ref{fig:perhist}.

% -----------------------------
\subsection{Herbig Ae/Be Stars}
\label{sec:herbig}
% -----------------------------

	Initially identified by \citet{herbig60}, Herbig Ae/Be stars are pre-main sequence (PMS) stars with masses $\gtrsim2~M_{\odot}$ and are the intermediate-mass analogs of T Tauri stars.  As a class their spectra display strong Balmer emission lines and they additionally exhibit significant infrared excess relative to the stellar photospheric continuum, attributed to emission from circumstellar gas and dust \citep{hernandez05, uemura04}.  Herbig Ae/Be stars are further characterized by aperiodic large-amplitude (up to 4 mag) variability with minima that initially redden and subsequently undergo a color \textquotedblleft turnaround\textquotedblright as the star becomes bluer as it gets fainter \citep{herbst99, grinin01}.  They are catagorized within the class of UXors (named after UX Orion), since it is thought that there is no intrinsic difference between the two \citep{herbst94, natta97, herbst99}.

\begin{figure}
\plotone{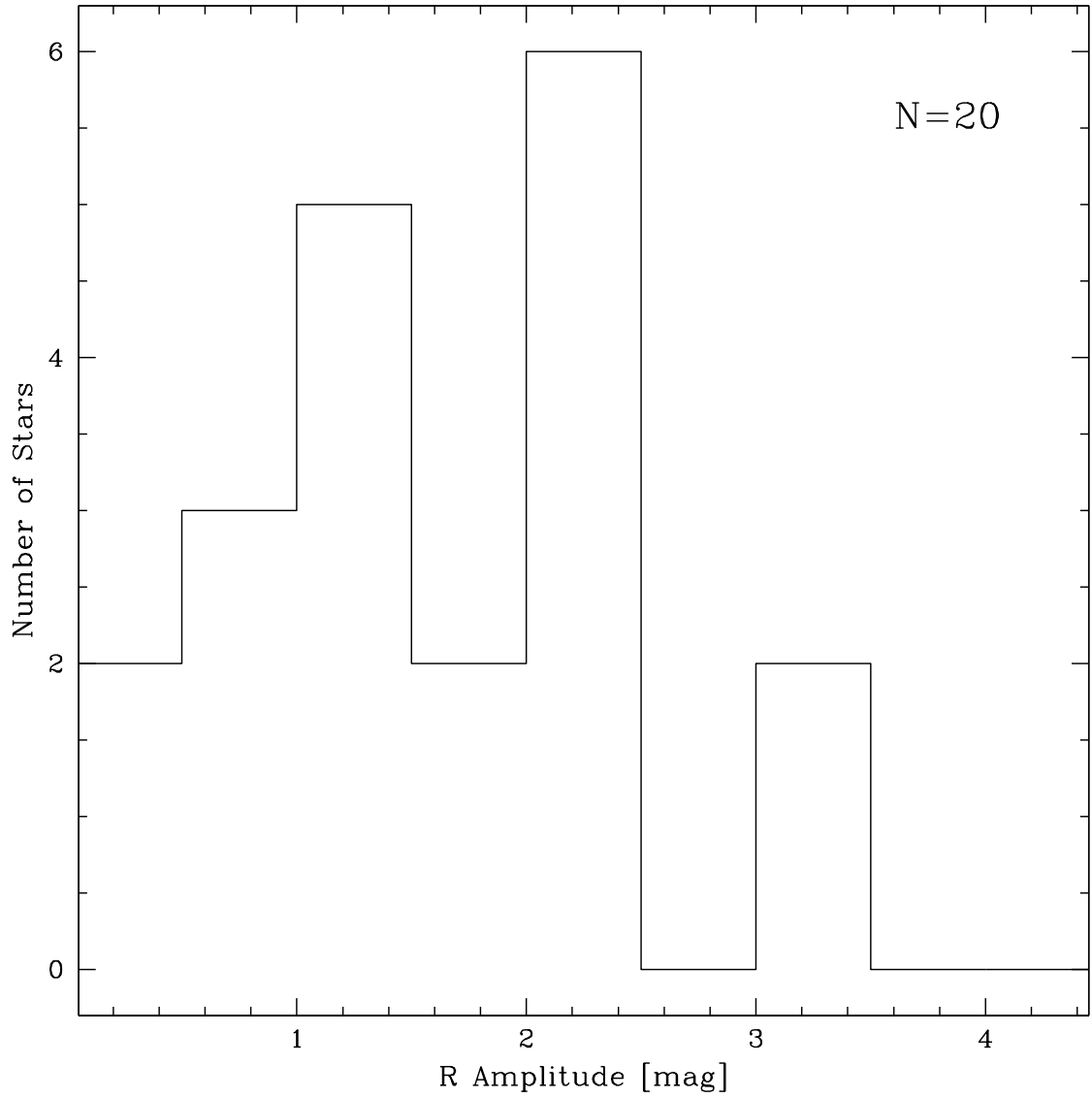}
\caption{A histogram of $R$-band amplitudes for the 20 potential Herbig Ae/Be stars.}
\label{fig:herbig.amphist}
\end{figure}

	We flag 20 of our variables as potential Herbig Ae/Be stars.  Eight of these show large-amplitude ($>$2.0 mag) variations while five of the remaining vary with lower amplitudes ($\lesssim$1.0 mag).  Figure~\ref{fig:herbig.amphist} shows a histogram of $R$-band magnitude ranges for all 20 potential Herbig Ae/Be stars.  The most interesting features are the dip in amplitude in the medium amplitude range (1.5--2.0 mag), the subsequent rise in amplitude (2.0--2.5 mag), and the presence of multiple (two) large-amplitude ($>$3.0 mag) variables.  Considered as a whole, the amplitude distribution for these potential Herbig Ae/Be stars in Cyg OB2 is in agreement with that found by \citet{herbst99}, including the three aforementioned features.

\begin{figure}
\plotone{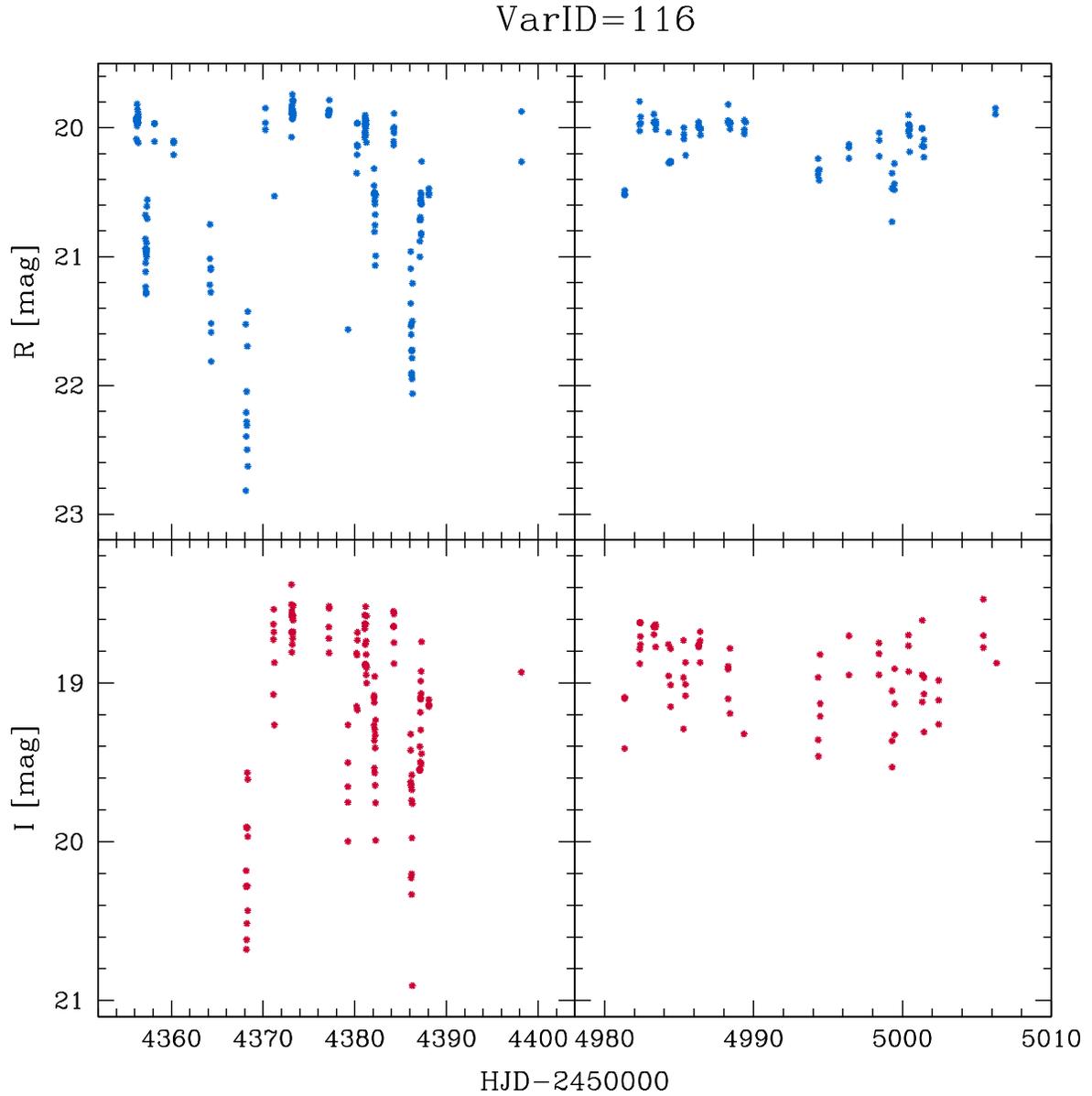}
\caption{The light curve of a potential Herbig Ae/Be star that is similar to that of MisV1147 \citep{uemura04}, exhibiting two distinct states.  One is a relatively calm state with lower amplitude ($\lesssim$0.5 mag) variations while the other is an active state with deep minima that can reach 3+ magnitudes.}
\label{fig:herbig116}
\end{figure}

\begin{figure}
\plotone{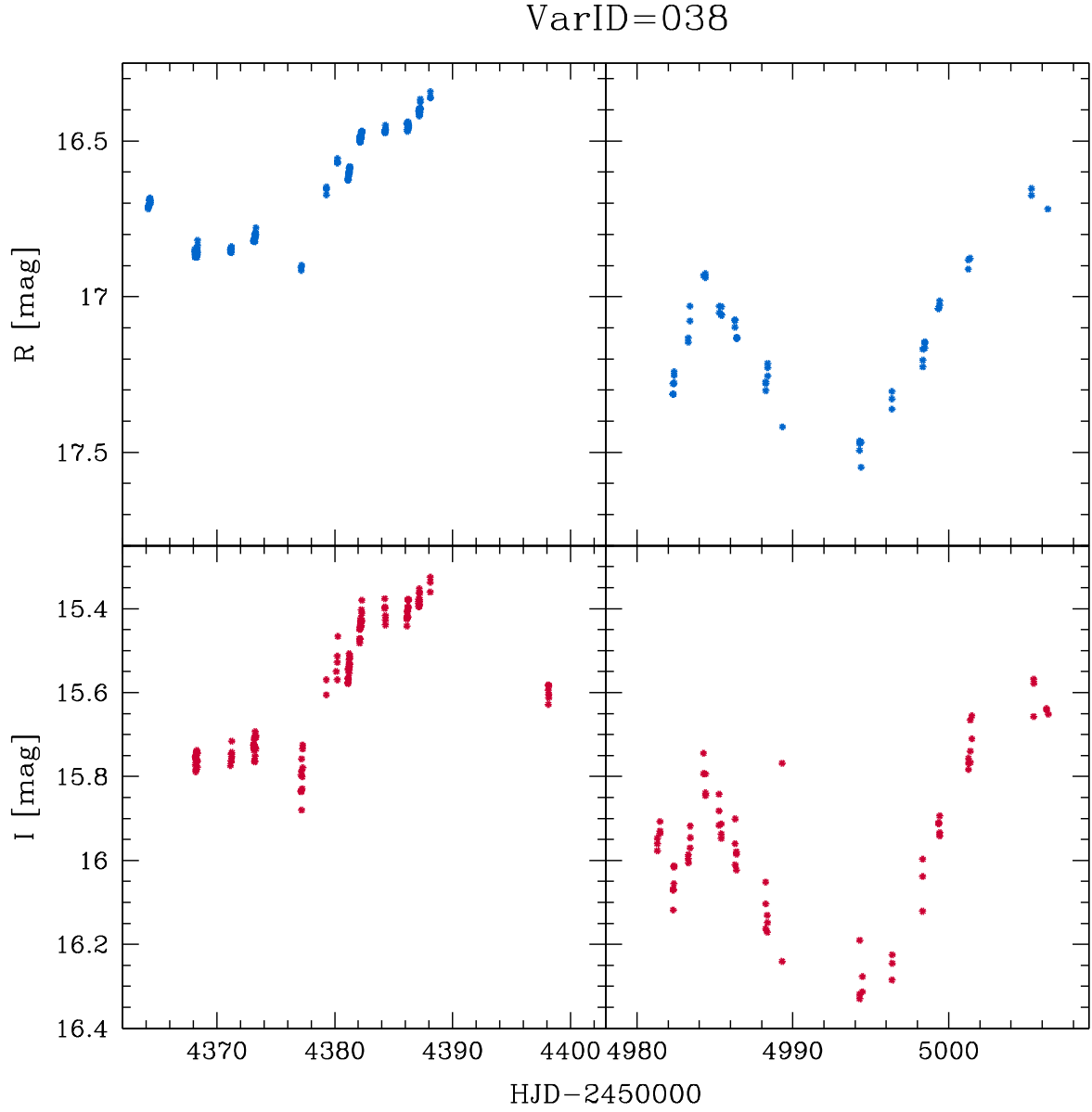}
\caption{The light curve of a candidate Herbig Ae/Be star that is similar to that of RR Tau \citep{herbst99}.  It is continuously active and spends roughly equal amounts of time at all magnitude levels.}
\label{fig:herbig038}
\end{figure}

	Several of our candidates display variations strikingly similar to known Herbig Ae/Be stars.  Four follow the pattern of MiSV1147 \citep{uemura04}, in which there are two distinct states.  The stars exhibit a calm state with variations of small ($\lesssim0.5$ mag) amplitudes and also experience fading episodes with deep minima of $\sim$1 (Varid 041), $\sim$2 (Varids 087, 100), and even $\sim$3+ (Varid 116) magnitudes.  Figure~\ref{fig:herbig116} shows a two-band light curve of such an object.  Many of the remaining candidates more closely resemble RR Tau, spending roughly equal amounts of time at different magnitude levels \citep{herbst99}.  Figure~\ref{fig:herbig038} shows an example light curve of a continuously active star.

	The easiest way to confirm these as Herbig Ae/Be stars would be to obtain spectra that display H$\alpha$ emission, as well as other Balmer emission lines.  Spectral classification that would identify these as K0 or earlier would at the least confirm them as intermediate-mass stars, though the lack of periodic signals that we find in their light curves is consistent with the results of \citet{herbst99} and perhaps indicates that these candidates similarly lack the surface hot or cool spots that are common among lower-mass PMS stars and that evince themselves as periodic modulations in the light curves.  A population of Herbig Ae/Be stars embedded within the central core of Cyg OB2 would help to constrain the age of the cluster.  Without further spectral information, however, the true status of these stars remains unconfirmed.

%%%%%%%%%%%%%%%%%%%%%%%%%%%%%%%%%%%%%%%%%%%%%%%%%%%%%%%%%%%%%%%%%%%%%%%%%%%%%%%
\section{Catalog Matching}
\label{sec:catmatch}
%%%%%%%%%%%%%%%%%%%%%%%%%%%%%%%%%%%%%%%%%%%%%%%%%%%%%%%%%%%%%%%%%%%%%%%%%%%%%%%

% -----------------------------
\subsection{2MASS Data}
\label{sec:2mass}
% -----------------------------

\begin{figure}
\plotone{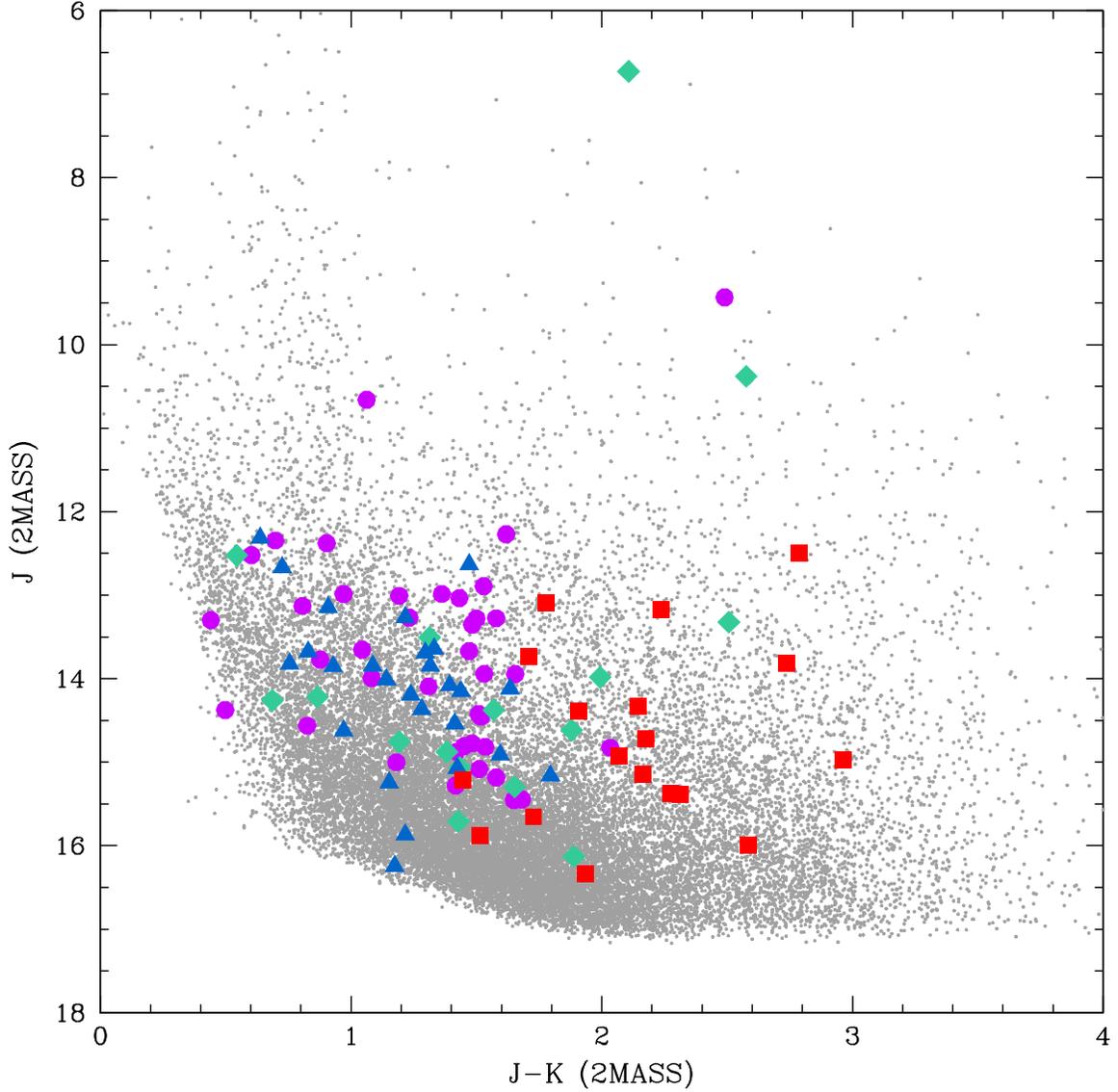}
\caption{A near-IR CMD for the 25,012 2MASS point sources within 30$\arcmin$ of the field center and with magnitude errors $<$0.2 mag.  Our 99 variables with 2MASS matches are overplotted in color.  Blue triangles represent eclipsing binaries and eclipsing binary candidates (26 with 2MASS matches), red squares indicate potential Herbig Ae/Be stars (18), purple circles mark pulsating variables (39), and light green diamonds denote all other variables (16).}
\label{fig:2masscmdvartyp}
\end{figure}

	Cyg OB2 suffers from high and variable extinction, $A_{V} \approx$ 5--20 magnitudes \citep{knodlseder00}, that is largely mitigated when observing in IR or NIR wavelengths.  We extract all 2MASS \citep{skrutskie06} point sources within 30$\arcmin$ of the center of our survey and remove any with errors $>$0.2 mag or a null value in any band (\textit{J}, \textit{H}, or \textit{K}), leaving 25,012 total sources.  Using a matching radius of 1$\arcsec$ we find unique matches to 99 ($\sim$82\%) stars in our variability catalog.  Figure~\ref{fig:2masscmdvartyp} shows a CMD for all 2MASS point sources remaining after the quality cut, with the different types of variability labeled in different shapes and colors.

% -----------------------------
\subsection{\textit{Chandra} Catalog}
\label{sec:chandra}
% -----------------------------

	The high extinction in the region can also be obviated with X-ray observations, as X-rays can penetrate extinction up to $A_{V}\approx500$ \citep{grosso05}.  The \textit{Chandra} catalog of \citet{wright09} contains 1,696 X-ray sources detected within Cyg OB2, with $<$1\% of false sources.  A total of 97 of our variables lie within one of their two \textit{Chandra} fields, and using a matching radius of 1$\arcsec$ we obtain unique matches to 49 ($\sim$51\%) of those 97 stars in our variability catalog.  They employ a one-dimensional Kolmogorov-Smirnov test to investigate X-ray variability, and of our 49 matched stars, seven display X-ray variability with three more being possibly variable.

\begin{figure}
\plotone{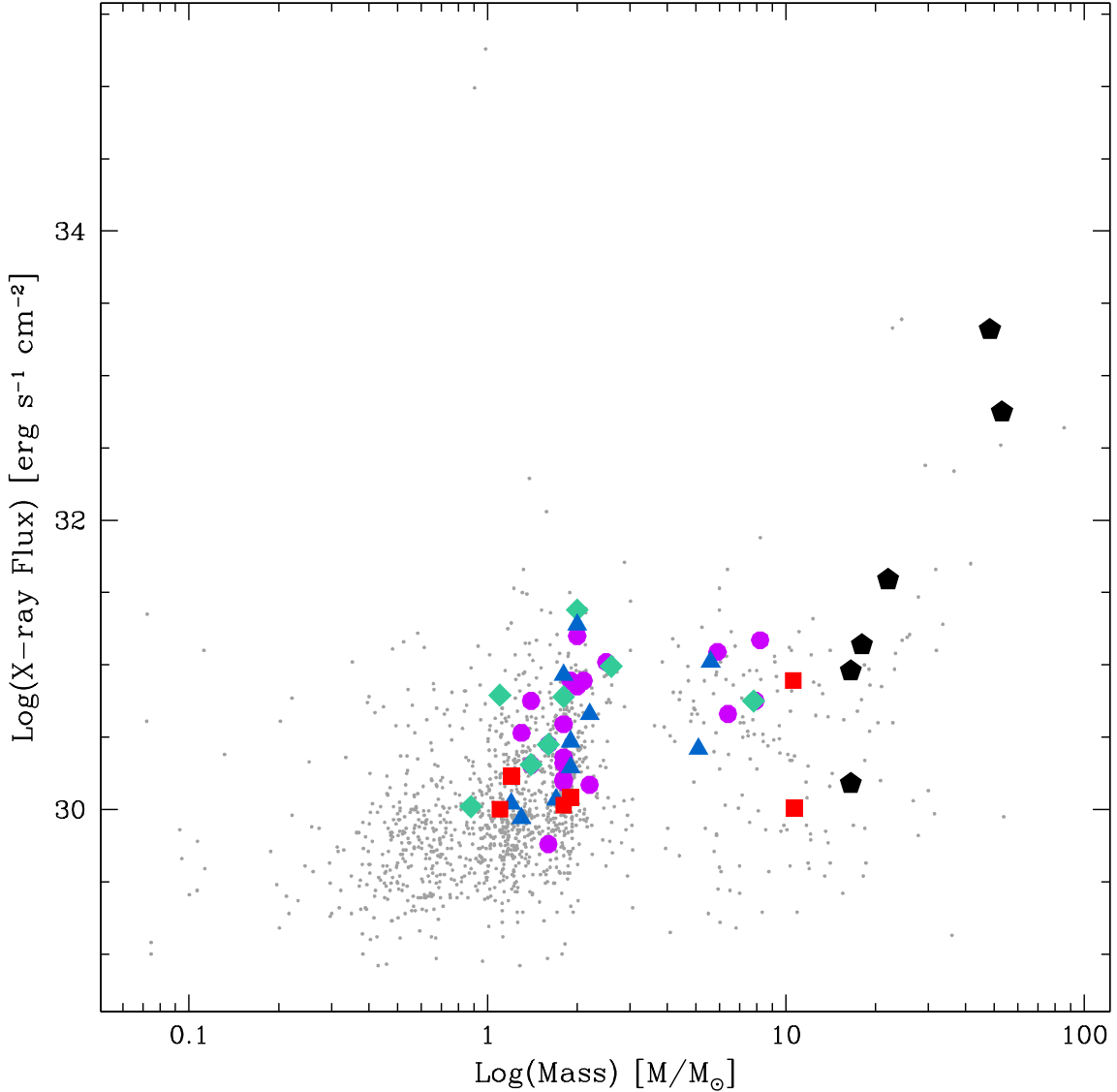}
\caption{X-ray flux as a function of stellar mass for the 1,417 sources from the catalog of \citet{wright09} that have stellar mass estimates.  Our 44 variables with X-ray matches and stellar mass estimates are overplotted in color.  Blue triangles represent eclipsing binaries and eclipsing binary candidates (10 with matches and mass estimates), red squares indicate potential Herbig Ae/Be stars (6), purple circles mark pulsating variables (20), and light green diamonds denote all other variables (8).  Six previously known OB binaries additionally have X-ray matches and are shown in black pentagons.}
\label{fig:xray}
\end{figure}

	They find optical and NIR counterparts for 1,501 of their sources.  Using a (\textit{J}, \textit{J}--\textit{H}) CMD they calculate individual stellar masses by tracing extinction vectors.  Figure~\ref{fig:xray} shows X-ray flux plotted as a function of the estimated stellar mass, with stellar variability classes overplotted in color.  It is clear that their mass estimates are reasonable for the six OB binaries, shown in purple, all of which have masses $>$15$~M_{\odot}$.  Eight of our EBs and EB-candidates that match and have mass estimates appear to be low-mass ($<$few~M$_{\odot}$), while two are of intermediate (2--8$~M_{\odot}$) mass.  The sample of our EBs and EB-candidates with mass estimates is incomplete, but such a diagnostic would be useful for constraining the binary fraction among high-mass early-type stars.

%%%%%%%%%%%%%%%%%%%%%%%%%%%%%%%%%%%%%%%%%%%%%%%%%%%%%%%%%%%%%%%%%%%%%%%%%%%%%%%
\section{Summary}
\label{sec:Summary}

	In this work we present a catalog of photometrically variable stars in the Cygnus OB2 association.  After generating \textit{R}- and \textit{I}-band light curves using ISIS, we run AoV and LS period searches and apply the \citet{stetson96} variability index, finding 121 stars with two-band variability, 116 of them new.  Of these, 99 have unique 2MASS NIR matches and 49 have \textit{Chandra} X-ray matches, with seven exhibiting X-ray variability.  We find 10 eclipsing binaries (EBs), 17 EB-candidates, and 20 potential Herbig Ae/Be stars, with most of the rest constituting an assortment of short and medium period pulsating variables.

	Follow-up work is underway to obtain spectra of the potential Herbig Ae/Be stars, as H$\alpha$ emission would be conclusive proof of their classification and help constrain the age of Cyg OB2.  Additionally, obtaining spectral types of the EBs and EB-candidates would be useful for constraining the binary fraction of massive stars.  Cyg OB2 has the largest number of spectroscopically identified O stars (currently 65) as well as a high concentration of as many as 2,400 OB stars \citep{knodlseder00}, making it an excellent target for finding massive binaries.  Additionally, accurate mass determinations of these potential early-type stars could add to the total number of known OB stars in Cyg OB2 and also provide evidence for or against the proposed propensity of massive binaries to form as \textquotedblleft twins\textquotedblright, with a mass ratio near unity \citep{pinsonneault06}.
%%%%%%%%%%%%%%%%%%%%%%%%%%%%%%%%%%%%%%%%%%%%%%%%%%%%%%%%%%%%%%%%%%%%%%%%%%%%%%%

%%%%%%%%%%%%%%%%%%%%%%%%%%%%%%%%%%%%%%%%%%%%%%%%%%%%%%%%%%%%%%%%%%%%%%%%%%%%%%%
\section*{Acknowledgments}

	We thank Jason Eastman for supplying the MDM 4K photometric reduction pipeline and for helping to modify it to correct for crosstalk contamination.  We thank John Beacom and Alceste Bonanos for their insight and stimulating discussion, Matt Kistler for his helpful comments, and Don Terndrup for his assistance with the observations.  CBH and KZS are supported in part by the NSF grant AST-0707982.  JLP acknowledges support from NASA through Hubble Fellowship grant HF-51261.01-A awarded by the STScI, which is operated by AURA, Inc., for NASA, under contract NAS 5-26555. 
%%%%%%%%%%%%%%%%%%%%%%%%%%%%%%%%%%%%%%%%%%%%%%%%%%%%%%%%%%%%%%%%%%%%%%%%%%%%%%%

%==============================================================================
%                                FIGURES
%==============================================================================

\begin{figure}
\plotone{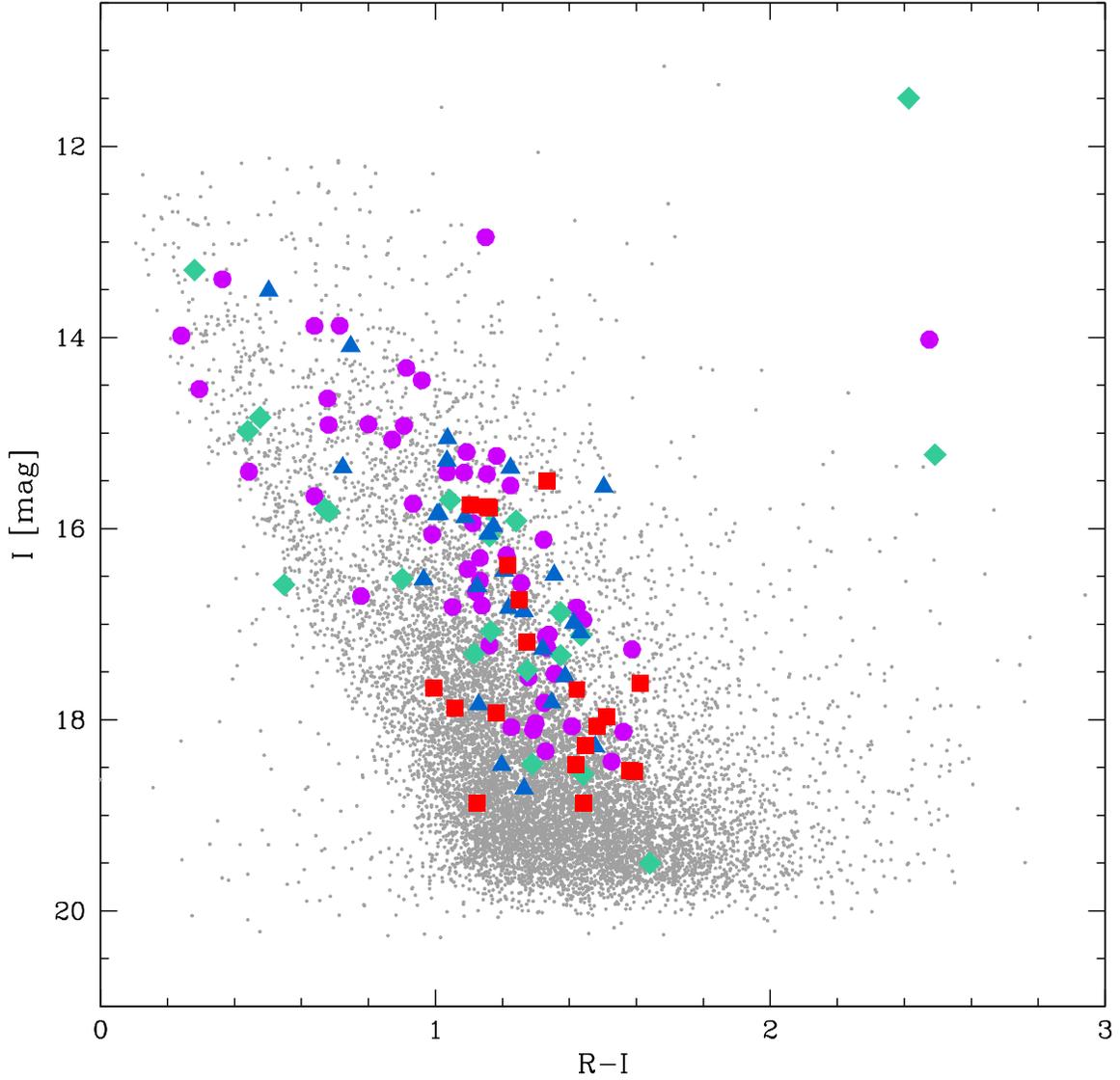}
\caption{An ($I$, $R$--$I$) CMD for all 11,690 stars with two-band photometry.  The different variability classes of our 121 photometric variables are overplotted in color.  Blue triangles represent eclipsing binaries and eclipsing binary candidates (27 in total), red squares indicate potential Herbig Ae/Be stars (20), purple circles mark pulsating variables (52), and light green diamonds denote all other variables (22).}
\label{fig:ricmd}
\end{figure}

%==============================================================================
%                                TABLES
%==============================================================================

\begin{deluxetable}{ccccccc}
\tabletypesize{\scriptsize}
\tablecaption{Summary of Observations \label{obs}}
\tablewidth{0pt}
\tablehead{
\colhead{Field} &
\colhead{$\alpha_{2000}$} &
\colhead{$\delta_{2000}$} &
\multicolumn{2}{c}{Exp Time [s]} &
\multicolumn{2}{c}{No. of Frames} \\
\colhead{} &
\colhead{} &
\colhead{} &
\colhead{R} &
\colhead{I} &
\colhead{R} &
\colhead{I}
}
\startdata
Field 1  &  20 33 16.91  &  +41 14 14.5  &  60  &  10  &  275  &  228  \\
Field 2  &  20 31 52.93  &  +41 27 00.4  &  60  &  10  &  293  &  232  \\
\enddata
\end{deluxetable}

\begin{deluxetable}{ccccccc}
\tabletypesize{\scriptsize}
\tablecaption{Previously known photometrically variable stars in Cyg OB2 \label{photvar}}
\tablewidth{0pt}
\tablehead{
\colhead{Lit ID} &
\colhead{Var Type} &
\colhead{Sp Type} &
\colhead{V} &
\colhead{Reference} &
\multicolumn{2}{c}{Matched?} \\
\colhead{} &
\colhead{} &
\colhead{} &
\colhead{[mag]} &
\colhead{} &
\colhead{R} &
\colhead{I}
}
\startdata
Schulte 7    &  Irr                &  O3If               &  10.55  &  1    &  S  &  S  \\
Schulte 18   &  Irr                &  BIIb               &  11.01  &  1    &  S  &  S  \\
Schulte 27   &  SB2/EW/KE          &  O9.5V              &  12.32  &  4    &  S  &  S  \\
Schulte 57   &  Unk                &                     &         &  1    &  V  &  V  \\
V0729 Cyg    &  EB                 &  O6.5-7 + early B?  &   9.21  &  2,3  &  S  &  S  \\
V2185 Cyg    &  EA EB              &  O9V + B9V-A0V      &  12.86  &  1    &  S  &  C  \\
V2186 Cyg    &  EA EB              &  B0V + ?            &  12.98  &  1    &  S  &  V  \\
V2187 Cyg    &  $\beta$ Ceph       &                     &  15.39  &  1    &  N  &  N  \\
V2188 Cyg    &  Be                 &  B1Ve-B3Ve          &  14.88  &  1,8  &  S  &  N  \\
V2189 Cyg    &  EA EB              &                     &  16.42  &  1    &  N  &  N  \\
V2190 Cyg    &  $\beta$ Ceph cand  &  B2Ve               &  14.06  &  1,8  &  C  &  C  \\
V2191 Cyg    &  EA EB              &  B4Ve               &  14.41  &  1,8  &  C  &  C  \\
MT91 512     &  Irr                &                     &  12.73  &  1    &  S  &  N  \\
MT91 390     &  Irr                &  O8V                &  12.95  &  1    &  S  &  C  \\
MT91 436     &  Unk                &                     &  16.08  &  1    &  N  &  N  \\
MT91 448     &  Unk                &  O6V                &  13.61  &  1    &  S  &  N  \\
MT91 456     &  Unk                &                     &  13.53  &  1    &  S  &  C  \\
MT91 460     &  Irr                &                     &  15.67  &  1    &  N  &  V  \\
MT91 503     &  Irr                &                     &  15.56  &  1    &  C  &  C  \\
MTE91 831    &  Irr                &                     &         &  1    &  V  &  V  \\
MTE91 849    &  EA EB              &                     &         &  1    &  V  &  V  \\
MTE91 876    &  Unk                &                     &         &  1    &  V  &  V  \\
MTE91 895    &  Irr/L              &                     &         &  1    &  N  &  N  \\
MTE91 900    &  Irr/L              &                     &         &  1    &  V  &  V  \\
MTE91 911    &  Irr/L              &                     &         &  1    &  N  &  N  \\
MTE91 916    &  Irr/L              &                     &         &  1    &  V  &  V  \\
MTE91 973    &  EB?                &                     &         &  1    &  C  &  V  \\
MTE91 974    &  Unk                &                     &         &  1    &  N  &  N  \\
MTE91 975    &  Unk                &                     &         &  1    &  N  &  N  \\
A36          &  SB2/EA             &  B0Ib + B0III       &  11.4   &  5-7  &  O  &  O  \\
\enddata
\tablecomments{O: outside our FoV; S: saturated in our data for that band; N: within our FoV and unsaturated but unmatched; C: matched but photometrically constant; V: matched and variable}
\tablerefs{
(1) \citet{pigulski98};
(2) \citet{rauw99};
(3) \citet{linder09};
(4) \citet{rios04};
(5) \citet{kiminki09};
(6) \citet{otero08a};
(7) \citet{hanson03};
(8) \citet{kiminki07}
}
\end{deluxetable}

\begin{deluxetable}{ccccccc}
\tabletypesize{\scriptsize}
\tablecaption{Spectroscopic binary stars in Cyg OB2 \label{specbin}}
\tablewidth{0pt}
\tablehead{
\colhead{Lit ID} &
\colhead{Bin Type} &
\colhead{Sp Type} &
\colhead{V} &
\colhead{Reference} &
\multicolumn{2}{c}{Matched?} \\
\colhead{} &
\colhead{} &
\colhead{} &
\colhead{[mag]} &
\colhead{} &
\colhead{R} &
\colhead{I}
}
\startdata
Schulte 1    &  SB1*           &  O8V                &  11.18  &  2     &  S  &  N  \\
Schulte 2    &  SB1*           &  B1I                &  10.64  &  1     &  S  &  N  \\
Schulte 3    &  SB2/EA?        &  O6IV? + O9III      &  10.35  &  3     &  O  &  O  \\
Schulte 8a   &  SB2            &  O5.5I + O6?        &   8.99  &  6,7   &  S  &  S  \\
Schulte 8c   &  SB1?           &  O5III              &  10.19  &  1     &  S  &  S  \\
Schulte 9    &  SB2            &  O5? + O6-7?        &  10.78  &  4     &  S  &  S  \\
Schulte 11   &  SB1*           &  O5I                &  10.03  &  1     &  O  &  O  \\
Schulte 15   &  SB1*           &  O8V + B            &  11.10  &  2     &  S  &  S  \\
Schulte 18   &  SB1*           &  B1I                &  11.01  &  1     &  S  &  S  \\
Schulte 20   &  SB1*           &  O9III + mid B      &  11.52  &  3     &  S  &  S  \\
Schulte 21   &  SB1*           &  B0Ib               &  11.42  &  1     &  S  &  S  \\
Schulte 26   &  SB1*           &  B1III              &  11.78  &  1     &  S  &  N  \\
Schulte 29   &  SB1?           &  O7V                &  11.91  &  1     &  O  &  O  \\
Schulte 41   &  SB1*/SB2?      &  B0V                &  13.49  &  1     &  S  &  C  \\
Schulte 73   &  SB2            &  O8III + O8III      &  12.5   &  3     &  O  &  O  \\
Schulte 74   &  SB1*           &  O8V                &  12.51  &  1     &  O  &  O  \\
V1393 Cyg    &  SB1*           &  B0Iab              &  11.07  &  1     &  S  &  S  \\
V1827 Cyg    &  SB2            &  O7? + O9?          &         &  9,10  &  O  &  O  \\
MT91 021     &  SB1*           &  B2II               &  13.74  &  1     &  O  &  O  \\
MT91 138     &  SB1*           &  O8I                &  12.26  &  1     &  S  &  S  \\
MT91 174     &  SB1*           &  B2IV               &  12.55  &  1     &  S  &  N  \\
MT91 196     &  SB1*           &  B6V                &  14.81  &  1     &  C  &  N  \\
MT91 202     &  SB1*           &  B2V                &  14.40  &  1     &  C  &  C  \\
MT91 234     &  SB1*           &  B2V                &  13.25  &  1     &  S  &  C  \\
MT91 238     &  SB1?           &  B1V                &  14.91  &  1     &  N  &  C  \\
MT91 241     &  SB1*           &  B2V                &  13.41  &  1     &  S  &  C  \\
MT91 252     &  SB2            &  B2III + B1V        &  14.15  &  2     &  C  &  C  \\
MT91 268     &  SB1*           &  B2.5V              &  14.38  &  1     &  C  &  C  \\
MT91 292     &  SB1*/SB2?      &  B2V                &  13.08  &  1     &  S  &  C  \\
MT91 298     &  SB1?           &  B3V                &  14.43  &  1     &  C  &  C  \\
MT91 311     &  SB1*           &  B2V                &  13.87  &  1     &  C  &  C  \\
MT91 336     &  SB1*           &  B3III              &  14.13  &  1     &  O  &  O  \\
MT91 365     &  SB2?           &  B1V                &  13.81  &  1     &  S  &  N  \\
MT91 372     &  EA*/SB1        &  B0V + B2?V         &  14.97  &  3,5   &  O  &  O  \\
MT91 403     &  SB1*           &  B1V                &  12.94  &  1     &  O  &  O  \\
MT91 428     &  SB1*           &  B1V                &  14.06  &  1     &  S  &  C  \\
MT91 448     &  SB1*           &  O6V                &  13.61  &  1     &  S  &  N  \\
MT91 485     &  SB1?           &  O8V                &  12.06  &  1     &  S  &  C  \\
MT91 492     &  SB1*           &  B1V                &  14.85  &  1     &  N  &  N  \\
MT91 493     &  SB1*           &  B5IV               &  14.99  &  1     &  O  &  O  \\
MT91 513     &  SB1*           &  B2V                &  14.26  &  1     &  S  &  C  \\
MT91 517     &  SB1?           &  B1V                &  13.74  &  1     &  S  &  N  \\
MT91 561     &  SB1*/SB2?      &  B2V                &  13.73  &  1     &  S  &  C  \\
MT91 573     &  SB1?           &  B3I                &  13.87  &  1     &  S  &  N  \\
MT91 605     &  SB2?           &  B1V                &  11.78  &  1     &  S  &  C  \\
MT91 620     &  SB1?           &  B0V                &  13.89  &  1     &  S  &  N  \\
MT91 639     &  SB2?           &  B2V                &  14.37  &  1     &  C  &  C  \\
MT91 692     &  SB2?           &  B0V                &  13.61  &  1     &  S  &  N  \\
MT91 720     &  SB2            &  early B + early B  &  13.59  &  2     &  O  &  O  \\
MT91 759     &  SB2?           &  B1V                &  14.65  &  1     &  O  &  O  \\
MT91 771     &  SB2            &  O7V + O9V          &  12.06  &  2     &  O  &  O  \\
A45          &  SB2            &  B0.5V + B2V?-B3V?  &         &  3,8   &  O  &  O  \\
\enddata
\tablecomments{SB1/2 denotes a confirmed spectroscopic binary, SB1/2* a probable binary, and SB1/2? a possible binary; O: outside our FoV; S: saturated in our data for that band; N: within our FoV and unsaturated but unmatched; C: matched but photometrically constant}
\tablerefs{
(1) \citet{kiminki07};
(2) \citet{kiminki08};
(3) \citet{kiminki09};
(4) \citet{naze08};
(5) \citet{wozniak04};
(6) \citet{romano69};
(7) \citet{debecker04};
(8) \citet{hanson03};
(9) \citet{stroud10};
(10) \citet{otero08b}
}
\end{deluxetable}

\begin{deluxetable}{cccccccccccc}
\rotate
\tabletypesize{\scriptsize}
\tablecaption{Parameters of the 121 variable stars identified in this study \label{varparams}}
\tablewidth{0pt}
\tablehead{
\colhead{VAR ID} &
\colhead{$\alpha_{2000}$} &
\colhead{$\delta_{2000}$} &
\colhead{Var Type} &
\colhead{2MASS ID} &
\colhead{R} &
\colhead{I} &
\colhead{J} &
\colhead{H} &
\colhead{K} &
\multicolumn{2}{c}{X-ray} \\
\colhead{} &
\colhead{} &
\colhead{} &
\colhead{} &
\colhead{} &
\colhead{[mag]} &
\colhead{[mag]} &
\colhead{[mag]} &
\colhead{[mag]} &
\colhead{[mag]} &
\colhead{Match} &
\colhead{Variable}
}
\startdata
001  &  20:31:00.18  &  41:34:30.07  &           &  20310017+4134300  &  13.58  &  13.29  &  12.523  &  12.089  &  11.979  &  O  &     \\
002  &  20:33:40.09  &  41:08:29.65  &  Pul      &                    &  13.76  &  13.39  &          &          &          &  N  &     \\
003  &  20:33:39.61  &  41:10:19.26  &           &  20333961+4110192  &  13.91  &  11.50  &   6.728  &   5.296  &   4.621  &  N  &     \\
004  &  20:33:42.13  &  41:22:22.77  &  EB       &  20334212+4122227  &  14.01  &  13.51  &  12.313  &  11.905  &  11.675  &  O  &     \\
005  &  20:34:04.04  &  41:14:43.04  &  Pul      &  20340404+4114430  &  14.10  &  12.95  &  10.660  &   9.999  &   9.598  &  O  &     \\
006  &  20:31:42.01  &  41:21:58.29  &  Pul      &  20314200+4121582  &  14.22  &  13.98  &  13.298  &  12.925  &  12.857  &  O  &     \\
007  &  20:32:19.94  &  41:33:55.08  &  Pul      &  20321994+4133550  &  14.52  &  13.88  &  12.520  &  12.122  &  11.917  &  Y  &  C  \\
008  &  20:33:56.32  &  41:18:17.07  &  Pul      &  20335631+4118170  &  14.59  &  13.88  &  12.345  &  11.928  &  11.647  &  N  &     \\
009  &  20:33:04.37  &  41:24:39.96  &  Pul      &                    &  14.84  &  14.54  &          &          &          &  O  &     \\
010  &  20:32:34.97  &  41:35:03.36  &  EB?      &  20323497+4135033  &  14.84  &  14.10  &  12.666  &  12.177  &  11.941  &  N  &     \\
011  &  20:33:31.45  &  41:20:57.62  &  Pul      &  20333144+4120576  &  15.23  &  14.32  &  12.270  &  11.424  &  10.651  &  N  &     \\
012  &  20:34:05.78  &  41:19:33.26  &           &                    &  15.31  &  14.84  &          &          &          &  N  &     \\
013  &  20:33:52.24  &  41:23:21.50  &  Pul      &                    &  15.32  &  14.64  &          &          &          &  O  &     \\
014  &  20:33:23.06  &  41:10:13.86  &  Pul      &  20332306+4110138  &  15.41  &  14.45  &  12.378  &  11.790  &  11.475  &  N  &     \\
015  &  20:32:09.99  &  41:18:22.87  &           &                    &  15.42  &  14.98  &          &          &          &  O  &     \\
016  &  20:31:23.59  &  41:29:48.97  &  Pul      &                    &  15.60  &  14.91  &          &          &          &  Y  &  C  \\
017  &  20:31:37.32  &  41:24:38.21  &  Pul      &  20313732+4124382  &  15.71  &  14.91  &  13.129  &  12.585  &  12.323  &  N  &     \\
018  &  20:33:39.86  &  41:21:45.19  &  Pul      &  20333985+4121451  &  15.83  &  14.93  &  12.988  &  12.384  &  12.017  &  N  &     \\
019  &  20:31:25.74  &  41:29:15.77  &  Pul      &  20312573+4129157  &  15.85  &  15.40  &  14.376  &  13.877  &  13.878  &  N  &     \\
020  &  20:33:08.73  &  41:18:51.55  &  Pul      &                    &  15.94  &  15.07  &          &          &          &  N  &     \\
021  &  20:33:08.33  &  41:15:43.26  &  EB       &  20330833+4115432  &  16.08  &  15.36  &  13.679  &  13.095  &  12.850  &  Y  &  C  \\
022  &  20:32:36.58  &  41:32:18.58  &  EB       &  20323658+4132185  &  16.09  &  15.05  &  13.148  &  12.533  &  12.239  &  Y  &  V  \\
023  &  20:32:17.79  &  41:20:00.85  &  Pul      &  20321779+4120008  &  16.29  &  15.20  &  12.891  &  11.838  &  11.362  &  O  &     \\
024  &  20:32:30.88  &  41:18:10.75  &  Pul      &  20323087+4118107  &  16.30  &  15.66  &  14.563  &  13.892  &  13.738  &  N  &     \\
025  &  20:32:37.86  &  41:28:52.60  &  EB?      &  20323786+4128525  &  16.32  &  15.29  &  13.263  &  12.361  &  12.047  &  Y  &  C  \\
026  &  20:32:29.87  &  41:32:10.16  &  Pul      &  20322986+4132101  &  16.42  &  15.24  &  12.986  &  12.010  &  11.624  &  Y  &  C  \\
027  &  20:33:31.78  &  41:20:47.35  &  Pul      &  20333177+4120473  &  16.45  &  15.41  &  13.270  &  12.361  &  12.038  &  Y  &  C  \\
028  &  20:32:41.11  &  41:24:12.53  &           &  20324111+4124125  &  16.46  &  15.79  &  14.221  &  13.557  &  13.355  &  N  &     \\
029  &  20:33:50.90  &  41:19:03.50  &  Pul      &  20335089+4119034  &  16.50  &  14.02  &   9.434  &   7.713  &   6.944  &  N  &     \\
030  &  20:32:25.91  &  41:35:58.92  &  Pul      &  20322590+4135589  &  16.50  &  15.41  &  13.279  &  12.282  &  11.779  &  Y  &  C  \\
031  &  20:33:47.13  &  41:13:16.59  &           &  20334713+4113165  &  16.51  &  15.82  &  14.259  &  13.734  &  13.573  &  Y  &  C  \\
032  &  20:33:21.96  &  41:11:40.91  &  Pul      &  20332196+4111409  &  16.58  &  15.43  &  13.008  &  12.204  &  11.816  &  N  &     \\
033  &  20:33:53.10  &  41:11:19.40  &  EB?      &  20335310+4111193  &  16.59  &  15.36  &  13.823  &  13.241  &  13.068  &  Y  &  C  \\
034  &  20:33:48.81  &  41:23:08.26  &  Pul      &  20334880+4123082  &  16.67  &  15.74  &  13.773  &  13.203  &  12.896  &  O  &     \\
035  &  20:33:15.59  &  41:21:33.58  &           &                    &  16.75  &  15.70  &          &          &          &  Y  &  C  \\
036  &  20:33:22.25  &  41:17:44.60  &  Pul      &  20332224+4117446  &  16.77  &  15.55  &  13.037  &  12.005  &  11.605  &  Y  &  V  \\
037  &  20:32:51.37  &  41:08:40.07  &  Her?     &  20325137+4108400  &  16.84  &  15.50  &  12.492  &  10.988  &   9.706  &  N  &     \\
038  &  20:33:08.60  &  41:19:15.41  &  Her?     &  20330859+4119154  &  16.86  &  15.76  &  13.168  &  11.970  &  10.932  &  Y  &  C  \\
039  &  20:32:23.76  &  41:29:13.03  &  EB?      &  20322376+4129130  &  16.86  &  15.85  &  13.843  &  13.053  &  12.757  &  Y  &  C  \\
040  &  20:32:30.17  &  41:34:49.27  &  EB       &  20323017+4134492  &  16.86  &  15.85  &  13.854  &  13.242  &  12.925  &  N  &     \\
041  &  20:32:23.88  &  41:20:59.79  &  Her?     &  20322387+4120597  &  16.93  &  15.77  &  13.738  &  12.699  &  12.028  &  N  &     \\
042  &  20:32:25.06  &  41:17:21.93  &  Her?     &  20322506+4117219  &  16.95  &  15.79  &  13.090  &  12.018  &  11.311  &  O  &     \\
043  &  20:32:10.88  &  41:22:05.57  &           &  20321087+4122055  &  16.96  &  15.80  &  13.506  &  12.597  &  12.193  &  N  &     \\
044  &  20:33:36.95  &  41:23:21.65  &  EB?      &  20333694+4123216  &  16.97  &  15.88  &  13.691  &  12.803  &  12.395  &  O  &     \\
045  &  20:32:26.23  &  41:29:49.51  &  Pul      &  20322622+4129495  &  17.05  &  16.06  &  13.995  &  13.188  &  12.913  &  Y  &  C  \\
046  &  20:33:18.00  &  41:10:20.67  &  Pul      &  20331800+4110206  &  17.06  &  15.94  &  13.652  &  12.909  &  12.608  &  N  &     \\
047  &  20:33:49.00  &  41:14:09.00  &  EB?      &  20334899+4114089  &  17.07  &  15.56  &  12.634  &  11.737  &  11.163  &  N  &     \\
048  &  20:33:01.78  &  41:11:10.77  &  CV       &                    &  17.14  &  16.59  &          &          &          &  Y  &  V  \\
049  &  20:32:19.92  &  41:21:11.68  &  EB?      &  20321992+4121116  &  17.15  &  15.98  &  13.644  &  12.677  &  12.313  &  N  &     \\
050  &  20:33:18.27  &  41:14:11.49  &           &  20331826+4114114  &  17.16  &  15.92  &  13.323  &  11.933  &  10.816  &  N  &     \\
051  &  20:32:11.73  &  41:26:12.21  &  EB?      &                    &  17.21  &  16.06  &          &          &          &  Y  &  C  \\
052  &  20:33:58.35  &  41:20:56.26  &           &                    &  17.24  &  16.08  &          &          &          &  Y  &  V  \\
053  &  20:31:53.13  &  41:33:31.20  &           &  20315312+4133311  &  17.43  &  16.53  &  14.754  &  13.875  &  13.561  &  Y  &  C  \\
054  &  20:33:08.18  &  41:06:26.55  &  Pul      &  20330817+4106265  &  17.44  &  16.12  &  13.278  &  12.118  &  11.699  &  N  &     \\
055  &  20:32:29.05  &  41:25:37.07  &  Pul      &  20322904+4125370  &  17.44  &  16.31  &  14.093  &  13.185  &  12.784  &  Y  &  C  \\
056  &  20:32:45.02  &  41:23:49.18  &  Pul      &                    &  17.48  &  16.71  &          &          &          &  N  &     \\
057  &  20:33:16.01  &  41:17:27.26  &  Pul      &  20331600+4117272  &  17.49  &  16.28  &  13.671  &  12.740  &  12.200  &  N  &     \\
058  &  20:32:08.03  &  41:27:14.22  &  EB       &  20320802+4127142  &  17.50  &  16.53  &  14.622  &  13.878  &  13.651  &  Y  &  C  \\
059  &  20:32:25.76  &  41:28:42.51  &  Pul      &  20322575+4128425  &  17.52  &  16.42  &  13.354  &  12.386  &  11.871  &  Y  &  C  \\
060  &  20:33:29.65  &  41:18:45.88  &  Her?     &  20332964+4118458  &  17.60  &  16.39  &  13.812  &  12.300  &  11.073  &  N  &     \\
061  &  20:33:40.81  &  41:21:14.65  &  EB?      &  20334081+4121146  &  17.65  &  16.44  &  14.076  &  13.117  &  12.683  &  Y  &  P  \\
062  &  20:32:20.31  &  41:23:23.81  &  Pul      &                    &  17.68  &  16.55  &          &          &          &  Y  &  C  \\
063  &  20:32:25.23  &  41:24:27.65  &           &  20322522+4124276  &  17.72  &  15.23  &  10.379  &   8.621  &   7.803  &  N  &     \\
064  &  20:32:07.49  &  41:24:35.49  &  EB?      &  20320748+4124354  &  17.73  &  16.61  &  14.014  &  13.125  &  12.873  &  Y  &  C  \\
065  &  20:33:27.54  &  41:20:15.88  &  Pul      &                    &  17.78  &  16.66  &          &          &          &  Y  &  C  \\
066  &  20:33:20.16  &  41:19:43.28  &  Pul      &  20332016+4119432  &  17.83  &  16.57  &  13.942  &  12.822  &  12.411  &  Y  &  C  \\
067  &  20:31:48.69  &  41:21:54.97  &  EB?      &  20314869+4121549  &  17.84  &  16.49  &  13.844  &  12.979  &  12.528  &  O  &     \\
068  &  20:32:01.92  &  41:31:36.74  &  Pul      &                    &  17.87  &  16.82  &          &          &          &  Y  &  C  \\
069  &  20:33:57.42  &  41:21:48.22  &  Pul      &                    &  17.95  &  16.81  &          &          &          &  O  &     \\
070  &  20:32:28.42  &  41:20:17.48  &  Her?     &  20322842+4120174  &  18.00  &  16.75  &  14.384  &  13.258  &  12.475  &  N  &     \\
071  &  20:32:26.34  &  41:24:09.54  &  EB?      &  20322633+4124095  &  18.05  &  16.83  &  14.540  &  13.588  &  13.126  &  N  &     \\
072  &  20:31:22.63  &  41:28:33.76  &  EB       &  20312263+4128337  &  18.13  &  16.87  &  14.192  &  13.368  &  12.954  &  N  &     \\
073  &  20:32:31.81  &  41:27:14.99  &           &                    &  18.24  &  17.08  &          &          &          &  Y  &  P  \\
074  &  20:33:41.86  &  41:10:20.35  &  Pul      &                    &  18.25  &  16.83  &          &          &          &  Y  &  V  \\
075  &  20:32:37.87  &  41:21:19.40  &           &  20323787+4121194  &  18.25  &  16.88  &  14.374  &  13.212  &  12.805  &  Y  &  C  \\
076  &  20:33:23.36  &  41:19:12.63  &  Pul      &  20332336+4119126  &  18.38  &  17.22  &  14.807  &  13.749  &  13.355  &  Y  &  C  \\
077  &  20:33:04.24  &  41:06:51.43  &  Pul      &  20330423+4106514  &  18.39  &  16.95  &  13.945  &  12.741  &  12.290  &  N  &     \\
078  &  20:33:57.05  &  41:06:54.89  &  EB       &  20335704+4106548  &  18.40  &  16.99  &  14.151  &  13.155  &  12.713  &  O  &     \\
079  &  20:33:28.11  &  41:21:40.26  &           &  20332810+4121402  &  18.43  &  17.31  &  14.876  &  13.899  &  13.494  &  N  &     \\
080  &  20:31:04.79  &  41:20:26.10  &  Pul      &  20310478+4120261  &  18.45  &  17.11  &  14.458  &  13.393  &  12.940  &  O  &     \\
081  &  20:33:26.34  &  41:12:38.85  &  Pul      &  20332633+4112388  &  18.46  &  17.13  &  14.423  &  13.380  &  12.914  &  Y  &  C  \\
082  &  20:31:53.46  &  41:23:48.71  &  Her?     &                    &  18.46  &  17.19  &          &          &          &  N  &     \\
083  &  20:32:53.77  &  41:15:13.40  &  EB?      &  20325377+4115134  &  18.52  &  17.09  &  14.122  &  12.961  &  12.488  &  Y  &  V  \\
084  &  20:33:12.73  &  41:11:33.02  &           &  20331273+4111330  &  18.55  &  17.12  &  13.983  &  12.674  &  11.989  &  Y  &  C  \\
085  &  20:34:08.62  &  41:06:59.75  &  EB       &  20340861+4106597  &  18.58  &  17.26  &  14.366  &  13.461  &  13.084  &  O  &     \\
086  &  20:32:31.04  &  41:18:55.09  &  Pul      &  20323104+4118550  &  18.58  &  17.25  &  14.817  &  13.713  &  13.282  &  N  &     \\
087  &  20:31:53.57  &  41:25:34.16  &  Her?     &  20315356+4125341  &  18.66  &  17.67  &  14.721  &  13.440  &  12.545  &  N  &     \\
088  &  20:32:32.93  &  41:18:23.14  &           &  20323292+4118231  &  18.70  &  17.33  &  14.617  &  13.422  &  12.737  &  N  &     \\
089  &  20:32:19.01  &  41:22:24.63  &           &  20321901+4122246  &  18.76  &  17.48  &  15.063  &  13.948  &  13.629  &  Y  &  C  \\
090  &  20:33:02.33  &  41:23:45.52  &  Pul      &  20330232+4123455  &  18.84  &  17.56  &  14.775  &  13.721  &  13.292  &  N  &     \\
091  &  20:32:23.90  &  41:16:48.49  &  Pul      &  20322389+4116484  &  18.85  &  17.26  &  15.002  &  14.018  &  13.821  &  O  &     \\
092  &  20:32:26.80  &  41:30:38.82  &  Pul      &  20322680+4130388  &  18.88  &  17.52  &  14.875  &  13.748  &  13.455  &  Y  &  C  \\
093  &  20:32:33.60  &  41:21:02.56  &  EB?      &  20323360+4121025  &  18.94  &  17.55  &  14.909  &  13.716  &  13.314  &  N  &     \\
094  &  20:31:16.89  &  41:30:01.23  &  Her?     &  20311688+4130012  &  18.94  &  17.89  &  15.651  &  14.498  &  13.924  &  O  &     \\
095  &  20:31:51.49  &  41:23:54.33  &  EB       &  20315149+4123543  &  18.97  &  17.84  &  15.243  &  14.371  &  14.091  &  N  &     \\
096  &  20:34:10.98  &  41:08:14.63  &  Her?     &  20341097+4108146  &  19.11  &  17.68  &  14.326  &  12.957  &  12.181  &  O  &     \\
097  &  20:33:04.50  &  41:24:16.25  &  Her?     &                    &  19.11  &  17.93  &          &          &          &  N  &     \\
098  &  20:33:09.93  &  41:17:56.39  &  Pul      &  20330992+4117563  &  19.15  &  17.82  &  14.827  &  13.538  &  12.793  &  Y  &  C  \\
099  &  20:31:59.75  &  41:27:23.16  &  EB?      &  20315974+4127231  &  19.17  &  17.82  &  15.069  &  13.971  &  13.645  &  Y  &  C  \\
100  &  20:33:18.89  &  41:14:37.15  &  Her?     &  20331888+4114371  &  19.23  &  17.62  &  14.971  &  13.170  &  12.009  &  Y  &  V  \\
101  &  20:31:18.74  &  41:34:55.14  &  Pul      &  20311873+4134551  &  19.30  &  18.08  &  15.279  &  14.208  &  13.862  &  O  &     \\
102  &  20:32:31.77  &  41:24:31.37  &  Pul      &  20323177+4124313  &  19.33  &  18.04  &  15.184  &  14.084  &  13.605  &  N  &     \\
103  &  20:32:55.97  &  41:22:34.33  &  Pul      &  20325597+4122343  &  19.40  &  18.11  &  15.452  &  14.283  &  13.802  &  N  &     \\
104  &  20:33:11.47  &  41:12:56.87  &  Pul      &                    &  19.48  &  18.07  &          &          &          &  Y  &  C  \\
105  &  20:33:55.21  &  41:20:01.19  &  Her?     &  20335520+4120011  &  19.49  &  17.97  &  14.929  &  13.659  &  12.862  &  N  &     \\
106  &  20:32:15.63  &  41:20:09.69  &  Her?     &  20321563+4120096  &  19.55  &  18.07  &  15.208  &  14.049  &  13.761  &  O  &     \\
107  &  20:31:47.13  &  41:26:36.18  &  Pul      &  20314713+4126361  &  19.66  &  18.33  &  15.081  &  14.022  &  13.569  &  Y  &  C  \\
108  &  20:31:30.27  &  41:28:49.52  &  EB?      &  20313027+4128495  &  19.67  &  18.47  &  15.866  &  14.850  &  14.650  &  N  &     \\
109  &  20:32:36.81  &  41:19:44.74  &  Pul      &  20323680+4119447  &  19.69  &  18.13  &  15.446  &  14.233  &  13.764  &  Y  &  C  \\
110  &  20:33:14.35  &  41:12:21.89  &  Her?     &  20331434+4112218  &  19.71  &  18.27  &  15.390  &  13.908  &  13.076  &  Y  &  C  \\
111  &  20:33:11.52  &  41:07:19.73  &           &  20331151+4107197  &  19.75  &  18.46  &  15.707  &  14.532  &  14.279  &  Y  &  C  \\
112  &  20:34:02.52  &  41:11:15.63  &  EB?      &  20340252+4111156  &  19.76  &  18.28  &  15.161  &  13.884  &  13.366  &  O  &     \\
113  &  20:33:08.69  &  41:21:10.25  &  Her?     &  20330868+4121102  &  19.89  &  18.47  &  15.994  &  14.376  &  13.409  &  N  &     \\
114  &  20:31:56.79  &  41:21:53.64  &  Pul      &                    &  19.96  &  18.44  &          &          &          &  O  &     \\
115  &  20:32:22.24  &  41:22:45.45  &  EB       &  20322224+4122454  &  19.99  &  18.72  &  16.243  &  15.127  &  15.069  &  N  &     \\
116  &  20:31:52.41  &  41:26:10.28  &  Her?     &  20315240+4126102  &  20.00  &  18.88  &  16.340  &  15.044  &  14.408  &  Y  &  C  \\
117  &  20:32:57.47  &  41:04:55.19  &           &  20325746+4104551  &  20.02  &  18.57  &  15.295  &  14.060  &  13.642  &  O  &     \\
118  &  20:33:12.84  &  41:14:40.91  &  Her?     &  20331283+4114409  &  20.11  &  18.53  &  15.378  &  13.799  &  13.104  &  N  &     \\
119  &  20:33:21.85  &  41:09:44.55  &  Her?     &  20332184+4109445  &  20.14  &  18.54  &  15.144  &  13.682  &  12.979  &  Y  &  C  \\
120  &  20:32:25.03  &  41:25:38.52  &  Her?     &  20322503+4125385  &  20.31  &  18.87  &  15.877  &  14.738  &  14.363  &  Y  &  C  \\
121  &  20:33:40.93  &  41:09:36.29  &           &  20334092+4109362  &  21.14  &  19.50  &  16.128  &  14.904  &  14.239  &  Y  &  P  \\
\enddata
\tablecomments{O: outside our FoV; N: within our FoV and unsaturated but unmatched; Y: matched; V: variable in X-rays; P: possibly variable; C: constant}
\end{deluxetable}

%==============================================================================
%                             BIBLIOGRAPHY
%==============================================================================

\end{document}